# Bragg Coherent Imaging of nanoprecipitates: role of superstructure reflections

Authors

**Maxime Dupraz[a,b]\*, Steven J. Leake[b] and Marie-Ingrid Richard[a,b]**

[a] IRIG MEM NRS, Univ. Grenoble Alpes, CEA Grenoble, 17 Avenue des Martyrs, Grenoble, 38000, France

[b] ESRF, 71 Avenue des Martyrs, Grenoble, 38000, France

Correspondence email: maxime.dupraz@cea.fr

**Funding information**    European Research Council (grant No. 818823); Agence Nationale de la Recherche (grant No. ANR-16-CE07-0028-01); Agence Nationale de la Recherche (grant No. ANR-18-ERC1-0010-01).

Coherent precipitation of ordered phases is responsible for providing exceptional high temperature mechanical properties in a wide range of compositionally complex alloys (CCAs). Ordered phases are also essential to enhance the magnetic or catalytic properties of alloyed nanoparticles. The present work aims at demonstrating the relevance of Bragg coherent diffraction imaging (BCDI) to study bulk and thin film samples or isolated nanoparticles containing coherent nanoprecipitates / ordered phases. Crystals of a few tens of nanometres are modelled with realistic interatomic potentials and relaxed after introduction of coherent ordered nanoprecipitates. Diffraction patterns from fundamental and superstructure reflections are calculated in the kinematic approximation and used as input to retrieve the strain fields using algorithmic inversion. We first tackle the case of single nanoprecipitates and show that the strain field distribution from the ordered phase is retrieved very accurately. Then, we investigate the influence of the order parameter $S$ on the strain field retrieved from the superstructure reflections and evidence that a very accurate strain distribution can be retrieved for partially ordered phases with large and inhomogeneous strains. In a subsequent section, we evaluate the relevance of BCDI for the study of systems containing many precipitates and demonstrate that the technique is relevant for such systems. Finally, we discuss the experimental feasibility of using BCDI to image ordered phases, in the light of the new possibilities offered by the 4[th] generation synchrotron sources.

## 1. Introduction

Precipitation strengthening with intermetallic compounds is the most effective approach for the enhancement of alloy strength in engineering structural materials, compared with solid-solution strengthening, grain-boundary strengthening, or work hardening (Wang *et al.*, 2018; Gladman, 1999). For instance, the coherent precipitation of ordered $L1_2$-$\gamma$' nanoprecipitates into a disordered face centred cubic (FCC)-$\gamma$ matrix confers to Nickel (Ni)-based superalloys an exceptionally high degree of strength, which is retained at high fractions of their melting point (Van Sluytman & Pollock, 2012). These unique properties are strongly linked to the microstructure of the Ni-based superalloys: due to the ordered nature of the $\gamma$'-phase, the dislocations are restricted to the $\gamma$-matrix up to a cutting stress. Thereby dislocations are confined in the $\gamma$-channels which constrains plastic deformation (Reed 2006). Coherent intermetallic precipitates, especially the $L1_2$-$\gamma$' ordered phase, have also demonstrated their efficiency for strengthening and improving thermal stability of Al-alloys (Knipling *et al.*, 2006; Wen *et al.*, 2013). Therefore, it is poignant to investigate precipitation in Al alloys to develop new light-weight materials that can be applied in high-temperature (HT) environments (>300 °C). In High-Entropy Alloys (HEAs), although initially a strong emphasis has been made on alloys with a single solid solution phase structure, the presence of an ordered phase in a solid solution matrix has been shown to be beneficial for obtaining a good combination of strength (including at high temperatures) and ductility (Senkov *et al.*, 2016; Stepanov *et al.*, 2019).

Coherent precipitation is not only the most efficient approach for the enhancement of alloys strength, but also the universal feature shared by all compositionally complex alloys (CCAs) with excellent HT mechanical properties. In particular, the perfect coherence between the ordered phase and the solid solution matrix is crucial for high temperature creep resistant properties of these alloys. In FCC alloys we have seen that the coherent ordered $L1_2$-$\gamma$' phase is responsible for the unique HT mechanical properties in many systems. It is observed for instance in the form of $Ni_3Al$ nanoprecipitates in Ni-based binary/ternary alloys (Giamei & Anton, 1985; Reed 2006; Kaufman *et al.*, 1989; Johnson & Voorhees, 1992), Ni/Co-based superalloys (Pyczak *et al.*, 2005; Ding *et al.*, 2018; Charpagne *et al.*, 2016) and HEA alloys (Shun & Du, 2009; Tong *et al.*, 2005) or $Al_3X$ (X= Sc, Zr, Er) nanoprecipitates in Al-alloys (Marquis & Seidman, 2001; Senkov *et al.*, 2008; Booth-Morrison *et al.*, 2011; Wen *et al.*, 2013; Clouet *et al.*, 2006). In other FCC alloys such as Al-Mg-Si (Andersen *et al.*, 1998; Klobes *et al.*, 2011) and Al-Cu (Biswas *et al.*, 2011) the maximum hardness is achieved in systems containing very fine (2.5 nm) fully coherent so-called Guinier-Preston zones (GP-I), sometimes in combination with semi-coherent, larger needles b'' (GP-II) zones (Andersen *et al.*, 1998). In body centred cubic (BCC) alloys on the other hand, precipitation strengthening is typically achieved by the introduction of the *B2* ordered phase. In Fe-based alloys, for instance (Ferritic steels), the *B2*-phase inserted in the form of coherent NiAl nanoprecipitates (Teng *et al.*, 2010; Vo *et al.*, 2014; Jiang *et al.*, 2017; Jiao *et al.*, 2015) confers HT creep-resistant properties of alloys due to the perfect coherence between the ordered phase and the solid solution matrix. The excellent mechanical properties of refractory HEAs (Senkov *et al.*, 2016; Jensen *et al.*, 2016) or Al-transition metals (TMs) – HEAs (Wang *et al.*, 2016; Ma *et al.*, 2017; Zhang *et al.*, 2018) can also be attributed to a superalloy-like microstructure: cuboidal *B2* NiAl or *L2$_1$*-Ni$_2$AlTi (Song *et al.*, 2017) nanoprecipitates coherently embedded in the BCC matrix.

Not only the volume fraction, size, and distribution of the precipitates but also their shape have a large impact on the mechanical properties of these alloys, in particular at elevated temperatures. In coherent precipitation, the nanoprecipitates size and shape are highly dependent on the lattice misfit between the ordered phase and solid-solution phase. Hence, in the case of coherent precipitation, control of the lattice misfit between the ordered phase and its parent solid solution is of vital importance to develop high-performance CCAs and improve the resistance to deformation (Wang *et al.*, 2014). In this optic, understanding the influence of the lattice misfit and the resulting elastic (coherency) strains in coherent nanoprecipitates is essential to understand their shape evolution during ageing. In addition, Cahn *et al.* (Cahn & Lärché, 1982) have shown that the shape of a coherent particle in a solid solution is not only controlled by the lattice misfit but also depends on the chemical interface. The particle equilibrium shape is therefore given theoretically by minimizing the sum of the interfacial energy $E_i$ and of the elastic energy $E_e$. The former scales with the surface while the latter scales with the volume and is therefore prevailing for larger precipitate sizes (Johnson & Voorhees, 1992; Voorhees, 1992; Thompson *et al.*, 1994). Therefore, many researchers have been exploring how to maintain the long-term stability of coherent precipitates through adjusting the amount of alloying elements or changing element species (Zhou, Ro *et al.*, 2004; Lo *et al.*, 2009; Zhou *et al.*, 2017). Ni-based alloys for instance typically possess a finite lattice misfit between the ordered $\gamma'$ precipitates and the disordered $\gamma$ matrix. Due to the elastic stresses associated with the misfit, precipitates undergo an evolution in shape during elevated temperature exposures, Ni$_3$Al precipitates can fission into smaller particles once they reach a critical size (Miyazaki *et al.*, 1982; Kaufman *et al.*, 1989; Glatzel & Feller-Kniepmeier, 1989) or change in shape with increasing particle size (Ardell & Nicholson, 1966; Doi *et al.*, 1985) in order to minimize their elastic energy. The difficulty to obtain cuboidal or spherical morphology of coherent *B2* or *L2$_1$* precipitates in BCC-based HEAs can also be explained by the large lattice misfit between the *B2* and BCC phases. In ferritic alloys on the other hand, the spherical shape of the NiAl precipitates, that is retained even for large precipitate size, indicates the dominance of the interfacial energy and low misfit strain between the matrix and the precipitate (Song *et al.*, 2015). In addition, several theoretical investigations (Ardell & Nicholson, 1966; Voorhees, 1992; Thompson *et al.*, 1994) and computer simulations (Abinandanan & Johnson, 1993; Wang & Mills, 1992; Goerler *et al.*, 2017) have indicated that shape evolution during coarsening is primarily controlled by the minimization of the elastic misfit strain ($E_e$) and of the interfacial energy ($E_i$) but can also be strongly modified and even impeded by the elastic interaction between misfitting precipitates. In order to design stable coherent microstructures in different solid-solution matrices it is of great interest to image the elastic strain field created in both; the nanoprecipitates and its matrix, and their evolution during ageing at elevated temperatures. This could give further insight for instance in the elastic interaction between the coherent precipitates and mobile dislocations which not only control the creep rate of these alloys (Larson, 1952) but is also related to the poor ductility of Ni-based alloys at room temperature (Semboshi *et al.*, 2019).

Coherent ordered precipitates are not only responsible for the unique mechanical properties of CCAs but can also have a large impact on the functional (magnetic, catalytic, …) properties of these alloys. For instance chemically ordered $L1_0$ alloyed nanoparticles such as FePt (Tzitzios *et al.*, 2005; Klemmer *et al.*, 2003), CoPt (Kitakami *et al.*, 2001; Klemmer *et al.*, 2002), NiPt (Cadeville *et al.*, 1986) and FePd (Oshima *et al.*, 1988; Klemmer *et al.*, 2002) attracted much attention for high density magnetic storage applications. In these alloyed nanoparticles, the transition from a disordered *A1* phase to an ordered $L1_0$ can enhance their magnetocrystalline anisotropy by several orders of magnitude (up to $10^7$– $10^8$ ergs/cm$^3$), the latter being intrinsic to the tetragonal symmetry of the $L1_0$ crystal structure. In isolated alloyed PtCu$_3$ nanoparticles an enhancement effect of structural ordering for the oxygen reduction reaction (ORR) was observed. Improved stability and enhanced activity were achieved in a partially ordered catalyst containing a disordered FCC core and a few nanometers thick $L1_2$ shell (Hodnik *et al.*, 2012).

Ideally, probing the defect state and elastic strain of individual precipitates requires high spatial resolution as well as high chemical sensitivity. So far, several techniques have been employed to characterise the crystal structure, composition, and crystal orientation of nanoprecipitates. Among them, transmission electron microscopy (TEM), (Bhat *et al.*, 1979; Vo *et al.*, 2014; Knipling *et al.*, 2008; Zhang *et al.*, 2018) atom probe tomography (APT) (Teng *et al.*, 2010; Jiang *et al.*, 2017; Xu *et al.*, 2015; Jiao *et al.*, 2014, 2015), electron backscatter diffraction (EBSD) or X-ray diffraction (XRD) (Wang *et al.*, 2016) are the most commonly employed. TEM and APT have the advantage of atomic resolution but are hindered by strong experimental constraints on the sample thickness and environment. These constraints are relaxed when using positron annihilation spectroscopy (PAS) in combination with *ab initio* calculations to refine the structures (Klobes *et al.*, 2011). This approach has a chemical sensitivity and allows to probe the defect state but is not sensitive to the elastic strain field. X-ray diffraction on the other hand shows several advantages: it is non-destructive, highly-strain sensitive, can penetrate a large amount of matter to probe embedded material and is sensitive to ordered phases. The latter feature is essential in order to characterize the strain and defect and strain states of ordered phases. In solid solution alloys, the atomic sites are randomly occupied by chemical elements. Long-range order is absent, and the scattering from such alloys is like the one from monoatomic crystals. In ordered phase alloys on the other hand, a given chemical element occupies one set of positions in the unit cell (for instance the corners) and the other element takes the other set of positions (for instance the cube centre positions). This long-range order gives rise to additional reflections in diffraction patterns known as superstructure reflections. These weak diffraction spots appear between the stronger fundamental reflections at the usual anti-Bragg positions (Warren, 1969) (**Figure 1**). In this work we aim at using these superstructure reflections in order to characterize the strain fields of coherent nanoprecipitates. To do so, a possible approach is to use Bragg Coherent Diffraction Imaging (BCDI): a lens less imaging technique that uses coherent X-rays in order to reconstruct real space images from the algorithmic inversion (Gerchberg, 1972; Fienup, 1982; Marchesini *et al.*, 2003) of high-resolution reciprocal space data (Robinson & Harder, 2009). In Bragg geometry the technique gives access to the displacement and strain fields of isolated crystals with a good spatial resolution (10 nm) and a very high strain sensitivity (few 10$^{-4}$) (Labat *et al.*, 2015; Cherukara *et al.*, 2018). It has a great potential for the study of coherent nanoprecipitates by imaging them using both fundamental and superstructure reflections and answer fundamental questions related to the interaction of dislocations with the elastic strain field of coherent precipitates.

A key limitation of BCDI is that it requires crystallographically isolated nano/micro-crystals in the range of 50 nm to 1 μm; large enough to give a strong scattering signal, but small enough to match the coherence volume of the X-ray beam. Only a small number of materials form crystals that fall into this size range, for example isolated metallic nanoparticles (Dupraz *et al.*, 2017) or metal thin films with grain size in this range (Yau *et al.*, 2017). The technique has therefore a great potential to image the nucleation of ordered phase using both fundamental and superstructure reflection in isolated crystals and fine-grained bulk samples. However, imaging of ordered precipitates embedded in a disordered matrix, is in principle not possible in single crystal bulk or thin film specimens. Using a FIB-based technique (Hofmann *et al.*, 2020), one can manufacture BCDI samples from bulk samples. However, in order to avoid a time consuming and delicate sample preparation, another possibility to image the shape and strain of precipitates is to measure a superstructure reflection that is only sensitive to the ordered phase. If the beam is larger than the precipitates it should be possible to use BCDI to reconstruct a single or an assembly of precipitates.

Here, we use numerical simulations to simulate precipitates embedded inside a matrix and to calculate the scattering from fundamental and superstructure reflections. In a first part, the case of a single coherent ordered precipitate is investigated in extensive details. The second part of the manuscript aims at quantifying the influence of the order parameter *S* on the strain retrieved from superstructure reflections. Finally, the last section aims to demonstrate the capability of Bragg Coherent Diffraction Imaging (BCDI) to retrieve from superstructure reflections the strain fields inside coherent nanoprecipitates.

## 2. Modelling / Tools & methods

The alloy considered in this study is a red gold alloy used in the jewellery industry (Plumlee, 2014). The manufacturing of red-gold components (nominal composition: Au, 43.3% at. Cu and 5.6% at. Ag) implies effectively controlling precipitation hardening. During its processing, this alloy typically undergoes a disorder-order transition from a chemically disordered FCC *A1* phase (space group $Fm\bar{3}m$) to an ordered tetragonal (FCT) *L1$_0$* phase (space group $P4/mmm$) (Garcia-Gonzalez *et al.*, 2019). The *L1$_0$*-ordered structure consists of alternate stacking of Au and Cu atoms along the *c*-axis of the FCT structure.

Two types of configurations are simulated in this study. Periodic boundary conditions are used for both types. The first type of simulation cells consists of spherical nanoprecipitates embedded in a parent matrix phase. The case of single precipitates is tackled in the initial part of the results section while assemblies of precipitates are considered in the last part of the same section. These nanoprecipitates ($L1_0$ phase) are coherent with the parent matrix ($A1$ phase), inducing large misfit strains. The orientation relationship between the nanoprecipitates and the matrix is well defined with three possible $L1_0$ variants: $L1_0$ [1 0 $\bar{1}$] ∥ $A1$ [1 $\bar{1}$ 0], $L1_0$ [1 $\bar{1}$ 0] ∥ $A1$ [1 $\bar{1}$ 0] and $L1_0$ [0 1 $\bar{1}$] ∥ $A1$ [1 $\bar{1}$ 0]. In this work, we mostly consider the $L1_0$ [1 0 $\bar{1}$] variant but the $L1_0$ [1 $\bar{1}$ 0] variant is also shown in *supporting information S1*. Unless otherwise specified, lattice orientations, which correspond with the axes of the simulation cell, are therefore $X_p$[1 0 $\bar{1}$], $Y_p$[1 0 1], $Z_p$[0 $\bar{1}$ 0] and $X_m$[1 $\bar{1}$ 0], $Y_m$[1 1 0], $Z_m$[0 0 1] for the $L1_0$ and $A1$ phases, respectively. This configuration is representative of the early stages of precipitation where both the size and spacing of the precipitates are of the order of a few nanometres (**Figure 2** & **Figure 3**). Semi-coherent and incoherent precipitates are also investigated in this work. The former are obtained by rotating the precipitates around the $Z_p$ axis. Above a critical angle, the elastic strain can no longer accommodate the lattice misfit, and interfacial dislocations are formed at the nanoprecipitate/matrix interface (**Figure S1**). In this case, the interface can be described as semi-coherent. Incoherent precipitates are inserted by further increasing the misorientation between the $L1_0$ and $A1$ phases (*supporting information S1*). Several crystal sizes are considered in this study. They range from 11 x 11 x 10.9 nm$^3$ to 66.1 x 66.1 x 65.8 nm$^3$ for the cells containing coherent nanoprecipitates, corresponding respectively to 89,600 and 19,360,100 atoms. The radii of nanoprecipitates range from 0.75 nm to 7.5 nm, which is consistent with the experimental observations (Garcia-Gonzalez *et al.*, 2019).

The second type of simulation cell in subsection 3.2. The latter contains only the tetragonal $L1_0$ AuCu phase and is used to characterize the influence of the long-range order parameter, $S$ (Warren, 1969). The simulation cell is shown in **Figure 4**a, the crystallographic directions corresponding with the axes of the simulation cell are $X_p$[1 $\bar{1}$ 0], $Y_p$[1 1 0], $Z_p$[0 0 1]. We use a 11.2 x 11.2 x 30.7 nm$^3$ simulation cell, which contains 267200 atoms. Note that these numbers can slightly fluctuate depending on the atomic composition of the alloy.

The interaction between atoms are modelled with two different embedded-atom model (EAM) (Daw & Baskes, 1983, 1984) potentials developed by (Foiles *et al.*, 1986) and (Zhou, Johnson *et al.*, 2004). It is well established that EAM potentials describe accurately the properties such as lattice and elastic constants, cohesive energies, and the vacancy formation energies of FCC metals. From these reliable monoatomic potentials, it is then possible to derive alloy potentials by specifically fitting the parameters of the potentials to alloy properties (such as the enthalpy of mixing). This approach has been employed to develop the EAM potentials used in this study.

Molecular Statics simulations are carried out with the open-source Large-scale Atomic/Molecular Massively Parallel Simulator (LAMMPS) (Plimpton, 1995). The system is relaxed at 0 Kelvin using a conjugate gradient algorithm. In this study, we used the Polak-Ribière version of this algorithm: at each gradient, the force gradient is combined with the previous iteration information to compute a new search direction perpendicular (or conjugate) to the previous search direction.

The three-dimensional (3D) diffraction patterns are calculated by summing the amplitudes scattered by each atom with its phase factor, following a kinematic approximation:

$$I(\boldsymbol{q}) = \left|\sum_j f_j(q) \cdot e^{-2\pi i \boldsymbol{q} \cdot \boldsymbol{r}_j}\right|^2 \tag{1}$$

where $\boldsymbol{q}$ is the scattering vector, $f_j(q)$ and $\boldsymbol{r}_j$ are respectively the atomic scattering factor and position of atom *j*. Note that the crystallographic convention is used in this manuscript, *i. e.* the $2\pi$ factor is not included in $\boldsymbol{q}$, which implies that a given *q* value corresponds to a real space distance *d* of *q* = 1/*d*. Equation (1) assumes fully coherent scattering. Absorption and refraction are not considered in this study since they are both negligible for simulation cells of few tens of nanometres.

Given the large number of atoms ($10^4 - 10^7$ atoms) and the similarly large number of points in the reciprocal space for which the calculation is performed ($10^5 - 10^7$ points), the computation is performed with a graphical processing unit (GPU) using the PyNX scattering package (Favre-Nicolin *et al.*, 2011). The calculations are carried out in the vicinity of Bragg positions $\boldsymbol{g}$ defined by their Miller indices *hkl*. Note that for all the reciprocal space points of a given reciprocal space map, we made the approximation that $f_j(q) = f_j(g)$. For the sake of simplicity, $\boldsymbol{g}$, which is a particular case of the generic scattering vector $\boldsymbol{q}$, will be referred as the scattering vector in the following.

Both fundamental and superstructure reflections are systematically calculated. We will see in the following sections that the former are generally sensitive to both phases, while the latter are only sensitive to the ordered $L1_0$ phase. This feature can be exploited to characterize the degree of ordering of a precipitate (subsection 3.2) and, very interestingly, can also be used to image an assembly of precipitates in an extended thin film or bulk specimen (subsection 3.3).

**Figure 1** shows slices of the reciprocal maps calculated from equation (1) for several configurations. The sensitivity of the superstructure reflections to the order parameter *S* is here obvious. The intensity of the superstructure reflections decreases with the order parameter (**Figures 1**d-f) and completely vanishes for *S* =0 (**Figure 1**f). The size and shape of the ordered phase are also reflected in the intensity distribution around the superstructure nodes. A single spherical $L1_0$ nanoprecipitate coherently embedded in a disordered *A1* matrix yields a spherical intensity distribution around the superstructure reflections (**Figures 1**b-c). For a fully ordered FCT $L1_0$ cell on the other hand, both fundamental and superstructure reflections reflect the tetragonal shape of the simulation cell (*Figure 1*a,d,e). The average strain values reported in section I and *supporting information S1* are obtained from the centre of mass of high order fundamental reflections (typically $\boldsymbol{g}$ = 0 0 8 and $\boldsymbol{g}$ = 0 0 10) calculated from equation (1). The real space displacement field, $\boldsymbol{u}(\boldsymbol{r})$, is calculated using two different methods. The first method relies on the calculation of the atomic scattering quantity, $\tilde{\rho}(\boldsymbol{r})$, directly from the complex scattered amplitude $\tilde{A}(\boldsymbol{q})$. $\tilde{\rho}(\boldsymbol{r})$ is also designated as a complex sample density/object, whose amplitude is the real Bragg electronic density, $\rho(\boldsymbol{r})$ and whose phase encodes the projection of the displacement field $\boldsymbol{u}(\boldsymbol{r})$ onto the scattering vector $\boldsymbol{g}$. The complex sample density can be obtained by performing a simple inverse Fourier transform (*FT$^{-1}$*) of the scattered amplitude:

$$\tilde{\rho}(\boldsymbol{r}) = \rho(\boldsymbol{r})e^{2\pi i \boldsymbol{g} \cdot \boldsymbol{u}(\boldsymbol{r})} = FT^{-1}(\tilde{A}(\boldsymbol{q})). \tag{2}$$

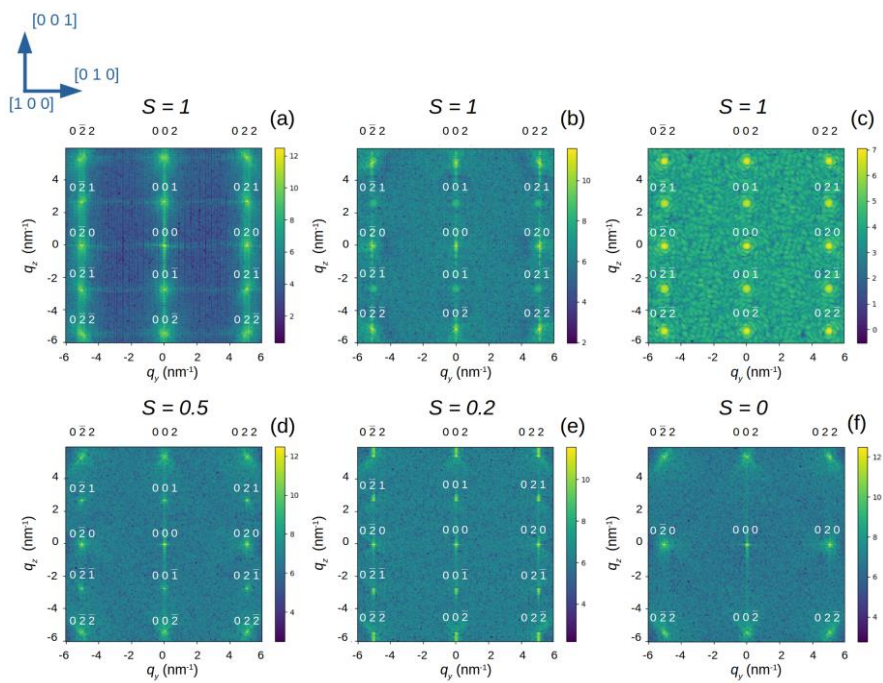

**Figure 1** *Slices of the reciprocal space intensity maps calculated for different atomistic configurations after relaxation* (a) Tetragonal simulation cell containing only a fully ordered $L1_0$ FCT phase ($S = 1$) (b) Single ordered $L1_0$ FCT nanoprecipitate ($r = 2.5$ nm) embedded into a disordered $A1$ FCC matrix (c) Same nanoprecipitate ($r = 2.5$ nm) without the disordered $A1$ phase. (d)-(f) Tetragonal simulation cell containing only a partially ordered $L1_0$ FCT phase: $S = 0.5$ (d), $S = 0.2$ (e), $S = 0$ (f). The reciprocal space volume (RSV) is kept constant in all figures and is equal to 11.99x11.99x12.01 nm$^{-3}$.

The projection of the displacement field $\boldsymbol{g}.\boldsymbol{u}(\boldsymbol{r})$ can then be compared to the atomic displacement field directly calculated from the relaxed atomic positions using OVITO (Stukowski, 2009). The strain field can then be derived from the displacement field: for instance, $\varepsilon_{xx} = \frac{\partial u_x}{\partial x}$.

If the complex sample density can be easily derived from the scattered amplitude in simulations, this is not the case experimentally, where a scattered intensity is measured on the far-field detector (square modulus of the scattered amplitude). However, if the diffraction pattern is sufficiently oversampled, the complex sample density may be reconstructed using phase retrieval algorithms (Miao *et al.*, 1999; Robinson & Harder, 2009). We aim at demonstrating that this BCDI-based approach is suitable to image the local displacement and strain fields inside and in the vicinity of coherent precipitates. This corresponds to the second method used in this paper to calculate the real space displacements.

The reconstruction of the samples is carried out using the PyNX CDI package (Mandula *et al.*, 2016). The initial support, *i.e.* the real space constraint, is estimated from the real autocorrelation of the diffraction intensity. A series of 1200 Relaxed Averaged Alternating Reflections (RAAR (Luke, 2004)) is followed by 200 Error-Reduction steps (Gerchberg, 1972; Fienup, 1982). In order to refine the support, the shrink-wrap algorithm is used every 20 iterations (Marchesini *et al.*, 2003). Note that the threshold is carefully adjusted for each dataset in order to maximize the success rate of the phase retrieval procedure (*supporting information S8*). For each simulated dataset 200 reconstructions with random phase start are performed. The quality of each reconstruction is then evaluated by comparing the retrieved displacement field component with the atomic displacement calculated from the relaxed atomic positions in OVITO. In order to allow a relevant comparison, the latter is averaged over a radius corresponding to the real space voxel size of the phase retrieval data. The quality of the reconstruction and in particular the accuracy of the displacement field is evaluated by comparing it to the calculations obtained from the atomistic positions (OVITO) and from the inverse Fourier transform of the scattered amplitude.

Successfully reconstructing the precipitate requires achieving a high spatial resolution, while satisfying the oversampling ratio, so that the phase retrieval algorithms can converge. The very small size of the spherical nanoprecipitates (ranging from 0.7 to 7.5 nm) requires a very high spatial resolution in the real space and therefore probing a large volume of the reciprocal space around the ***g*** of interest, significantly larger than the one typically measured experimentally (Labat *et al.*, 2015; Dupraz *et al.*, 2017). This allows generally to retrieve simultaneously the *A1* and *L1$_0$* phases from a fundamental reflection. Indeed, despite the large lattice mismatch between both phases, both can be retrieved accurately if the Bragg peaks are included in the calculated reciprocal space volume (RSV) (*supporting information S2*). Typically, the scattering is computed on 120x120x120 reciprocal space points (RSPs) with a varying sampling partially controlled by the size of the simulation cell. For the largest samples (66x66x66 nm$^3$) for instance, the sampling is varied between 1/900 reciprocal lattice units (r.l.u., corresponding to 2.84x10$^{-3}$ nm$^{-1}$ for the FCC reference lattice, a = 3.901 Å) and 1/240 r.l.u. (1.05x10$^{-2}$ nm$^{-1}$). The reciprocal space volume (RSV) measured is thus equal to 0.34x0.34x0.34 nm$^{-3}$ for the fine sampling and 1.25x1.25x1.25 nm$^{-3}$ for the coarse sampling. This translates to real space voxel sizes of 2.95 nm and 0.731 nm and oversampling ratios of 119 and 2.65 respectively. The latter is still fine enough to fulfil the oversampling criterion (s > 2) defined by (Miao *et al.*, 1998)

For the smallest sample (11x11x11 nm$^3$) (subsections 3.1 and 3.3), the Fourier space is probed with coarser steps (typically 1/80 r.l.u. or 3.2 x10$^{-2}$ nm$^{-1}$). If the scattering is computed on 140x140x140 RSPs, this gives a RSV of 4.50x4.50x4.50 nm$^{-3}$, which corresponds to a voxel size of 0.222 nm. Note that this value is smaller than the first neighbour distance in both the *A1* and *L1$_0$* phases (around 0.276 nm).

For the case of the FCT *L1$_0$* configuration (11x11x31 nm$^3$) (subsection 3.2), the sampling is kept constant (1/150 r.l.u. or 1.71 x10$^{-2}$ nm$^{-1}$ corresponding to s = 46) but the RSV is varied between 0.827x0.827x1.19 nm$^{-3}$ and 2.51x2.51x3.64nm$^{-3}$ corresponding to 50x50x66 and 150x150x200 RSPs respectively. These RSVs translate to an average voxel size of 1.09 nm and 0.357 nm, respectively.

The integrated amplitudes and intensities are calculated by summing the intensities of each RSP over RSVs of 0.827x0.827x1.19nm$^{-3}$ (50x50x66 RSPs) or smaller 0.413x0.413x0.593nm$^{-3}$ (25x25x33 RSPs), centred around the centre of mass of the considered reciprocal space nodes. This range was selected because of the excellent agreement with the theoretical values given by *Eqs*. (8) and (9). Further increasing the RSV for the integration tends to worsen the agreement with the theory (*supporting information S4-S6*). The integrated electron densities ($\rho_{int}$) are calculated by summing the amplitude of each voxel in the real space image. Only the voxels whose density is larger than an arbitrary fraction of the maximum of electron density in the crystal are considered for the calculation. Therefore, varying the threshold can affect the value of $\rho_{int}$, in particular for a small voxel size and can influence the value of $\rho_{int}$ for the superstructure reflections (*supporting in formation S5*). In this work, the threshold is typically set between 25% and 40% of the maximum electron density for the small RSV and between 12.5% and 25% for the large RSV. Finally, the accuracy of the strain fields retrieved from the superstructure reflections is assessed by calculating the Pearson correlation coefficients of the latter with the strain fields retrieved from the fundamental reflections that probe the same strain component.

## 3. Results

### 3.1. Imaging of a single coherent nanoprecipitate

In this first section, we consider the case of a single $L1_0$ precipitate coherently embedded in the disordered $A1$ matrix phase. **Figure 2**a shows a slice taken at the centre of the simulation cell containing a nanoprecipitate of radius r = 1.755 nm. The slice is oriented along the [0 $\bar{1}$ 0] direction of the $A1$ matrix phase. In the initial configuration, the nanoprecipitate is inserted with two distincts crystal structures: in the first case hereafter referred as the FCC configuration, a = c = 3.901 Å, *i. e.* the average lattice parameter of the solid solution ($A1$ phase) and in the second case, hereafter referred as the FCT configuration, a = 3.95 Å and c = 3.65 Å. This case gives a $\frac{c}{a}$ ratio, also known as the degree of tetragonality, of 0.92. This value corresponds to the equilibrium value in undeformed bulk specimens reported in the literature (Volkov, 2004). **Figure 2**b shows the *z*-component of the atomic strain tensor ($\varepsilon_{zz}$) after energy minimization for the FCT configuration. This direction corresponds to the [0 0 1] crystallographic direction of the $A1$ matrix phase ($\varepsilon_{zz} = \varepsilon_{001}$) and is aligned with the *c*-axis of the $L1_0$ nanoprecipitate (*i.e.* the [0 $\bar{1}$ 0] crystallographic direction). In the following $\varepsilon_{001}$ always refers to the strain component along the *c*-axis of both $L1_0$ (nanoprecipitate) and $A_1$ phases. As seen from **Figure 2**b, an homogeneous tensile strain builds up during relaxation in the nanoprecipitate region ($\overline{\varepsilon_{001}}$ = +4.1%), associated to an increase of the $\frac{c}{a}$ ratio from 0.92 to 0.96. The magnitude of this tensile strain is in reasonably good agreement with the experimental values measured by X-ray diffraction (Garcia-Gonzalez *et al.*, 2019; Garcia-Gonzalez *et al.*, 2020), where $\overline{\varepsilon_{001}}$ reaches a maximum value of 3.5% *(supporting information S11)*. In the surrounding $A1$ matrix phase, alternating regions of compressive and tensile strain are observed. These strain fluctuations are associated to local variations of the atomic composition, since the atoms are distributed randomly on the FCC lattice sites. The average lattice parameter of the $A1$ matrix phase remains constant during the relaxation, therefore the average strain value in the matrix is close to zero ($\overline{\varepsilon_{001}} \approx 0$). In contrast, **Figure 2**c shows the $\varepsilon_{001}$ atomic strain component after relaxation for the FCC configuration. The strain distribution in the *A1* matrix phase is identical to the FCT configuration, except in the close vicinity of the precipitate; in both cases the initial reference lattice is the disordered *A1* FCC phase which results in the same strain distribution after relaxation. On the other hand, the strain distribution in the nanoprecipitate after relaxation differs significantly from the FCT configuration. Interestingly, in contrast with the FCT configuration, a compressive strain builds up during relaxation ($\overline{\varepsilon_{001}} \approx -2.6\%$). The latter corresponds to a decrease of the $\frac{c}{a}$ ratio from 1 to 0.96 and illustrates the importance of the choice of the reference lattice for the calculation of the atomic strain. At first glance, one could indeed conclude that the relaxed strain state of the FCC nanoprecipitate differs from the FCT nanoprecipitate, however, this interpretation is not correct as revealed by the computation of $\varepsilon_{001}$ strain maps from *Eqs.* (1-2) for the FCC and FCT configurations (**Figures 2**d-i). **Figures 2**d-e show the $\varepsilon_{001}$ component for the FCC and FCT configurations, respectively, before energy minimization. The real space voxel size is set to 0.225 nm by tuning the size of the RSV. For both configurations, the $\varepsilon_{001}$ strain component is retrieved from the fundamental Bragg reflection (*g* = 0 0 2) and the same reference FCT lattice (a = 3.95 Å and c = 3.65 Å) is used to calculate $\varepsilon_{001}$. As illustrated in **Figures S3**c,d, using the FCT reference implies that the RSV is centred around the intensity scattered by the spherical nanoprecipitate. Since the *c*-parameter of the *A1* FCC phase is much larger than the reference *c*-parameter of the FCT reference lattice (3.901 Å vs 3.65 Å), a large tensile strain phase ($\overline{\varepsilon_{001}} \approx 6.9\%$) is observed in both cases in the *A1* matrix. The latter is consistent with the lattice mismatch between the *c*-axes of the FCT and FCC lattices. In the precipitate region, the strain state depends on the crystal structure of the precipitate. The FCC - precipitate, whose *c*-parameter is the same as the surrounding matrix, exhibits the same high tensile strain as the matrix (6.9%, **Figure 2**d). On the other hand, the lattice parameter of the FCT precipitate is set to the reference FCT lattice, therefore $\overline{\varepsilon_{001}} = 0$ in the initial configuration (**Figure 2**e). Using a superstructure reflection (*g* = 0 0 1) on the same configuration leads to similar observations in the FCT *L1₀* precipitate region ($\overline{\varepsilon_{001}} = 0$, **Figure 2**f). On the other hand, the scattered amplitude ($\tilde{A}(q)$) from the disordered *A₁* matrix is very weak. Hence $\tilde{\rho}(r)$, the inverse Fourier transform of $\tilde{A}(q)$, is also very weak and the strain is not retrieved accurately in the matrix region, whereas it is well retrieved in the precipitate region. **Figure 2**g-i show the $\varepsilon_{001}$ strain distribution for the same configurations after energy minimization. Again the same FCT reference lattice is used for both FCC and FCT configurations.

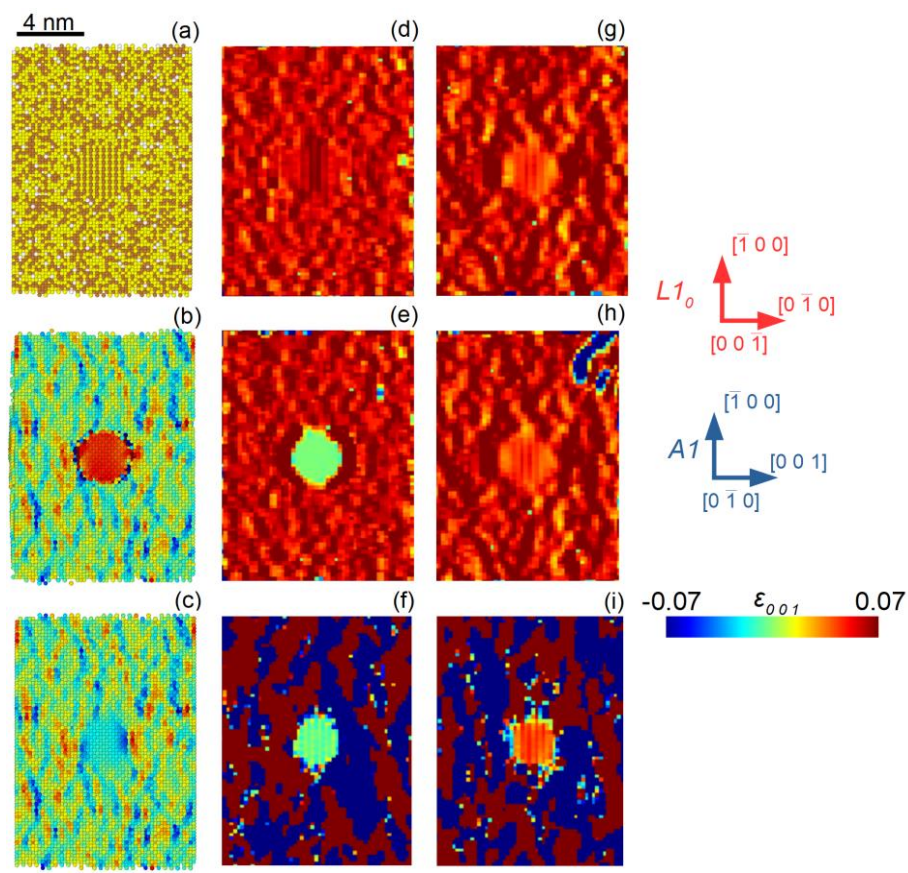

**Figure 2** (a) Atomistic configuration showing a *L1₀* nanoprecipitate coherently embedded in the A1 parent matrix The brown, yellow and silver atoms correspond to Cu, Au and Ag, respectively. (b)-(c) $\varepsilon_{001}$ atomic strain component after energy minimization (relaxation) of a simulation cell containing a FCT (FCC) precipitate respectively. (d)-(e) $\varepsilon_{001}$ retrieved from equation (2) for an unrelaxed configuration containing a FCC and FCT precipitate, respectively. Both strain maps are calculated using a fundamental reflection $g$ = 0 0 2. (f) $\varepsilon_{001}$ before relaxation for the FCT precipitate. The strain map is retrieved from a superstructure reflection $g$ = 0 0 1. (g)-(h) $\varepsilon_{001}$ after relaxation for the FCC and FCT precipitates, respectively, using $g$ = 0 0 2. (i) $\varepsilon_{001}$ after relaxation for the FCT precipitate using $g$ = 0 0 1. The strain maps shown in panels (d)-(i) are calculated from equation (2) and use the perfect FCT lattice (a = 3.95 Å and c = 3.65 Å) as a reference for the calculation. For all strain maps, the calculations are performed on a RSV of 4.50x4.50x4.50 nm⁻³, which corresponds to a voxel size of 0.222 nm.

Interestingly, and in contrast to the calculations from the relaxed atomic positions, the strain distribution is not only identical in the *A1* matrix region but also in the *L1₀* nanoprecipitates (**Figures 2**g,h). For $g$ = 0 0 2 average strain values of $\overline{\varepsilon_{001}} \approx$ +4.1% and $\overline{\varepsilon_{001}} \approx$ 6.9% are found for both configurations in the *L1₀* and *A1* phases, respectively. Both the magnitude and strain distribution are consistent with the one calculated directly from the relaxed atomic position of the FCT configuration. In the *A1* matrix region, the average *c*-parameter remains constant and equal to the *c*-parameter of the FCC lattice, explaining the large tensile strain with respect to the reference FCT lattice. For the superstructure reflection ($g$ = 0 0 1), the calculation of $\overline{\varepsilon_{001}}$ gives the same value of +4.1% (**Figure 2**i) with a strain distribution very consistent with the one obtained from the fundamental reflection. Similarly to the unrelaxed state (**Figure 2**f), the strain is not retrieved accurately in the matrix region.

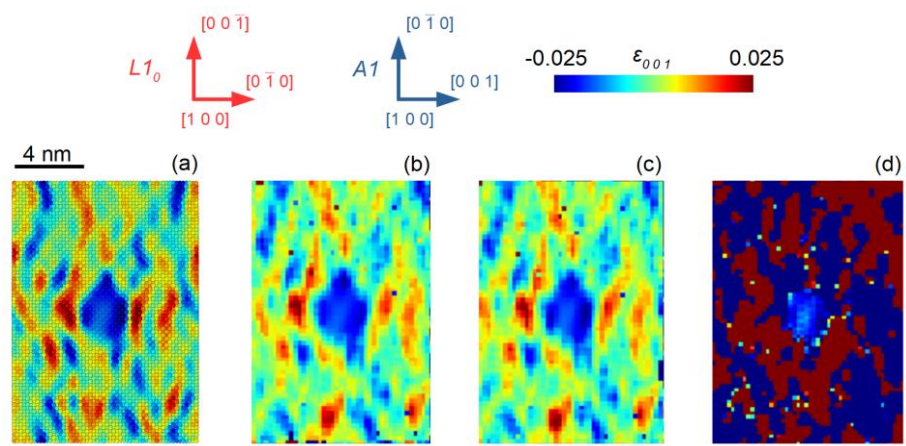

**Figure 3** (a) $\varepsilon_{001}$ atomic strain component after energy minimization of a simulation cell containing a FCC nanoprecipitate. (b)-(c) $\varepsilon_{001}$ after relaxation for the FCC and FCT precipitates respectively using *g* = 0 0 2. (d) $\varepsilon_{001}$ after relaxation for the FCT precipitates using *g* = 0 0 1. The strain maps shown in panels (b)-(d) are calculated from equation (2) and use the perfect FCC lattice (a = 3.901 Å and c = 3.901) as a reference for the calculation. For all strain maps, the calculations are performed on a RSV of 4.50x4.50x4.50 nm$^{-3}$, which corresponds to a voxel size of 0.222 nm.

**Figure 3**a displays $\varepsilon_{001}$ after energy minimization for the FCC configuration. The slice is oriented along the [1 0 0] crystallographic direction of the *A1* matrix phase. Note that in order to obtain a smoother representation of the atomic strain, the latter is averaged over a radius of 0.7 nm much larger than the 0.3 nm cut-off radius used in **Figures 2**b-c. As seen in **Figure 3**a, this larger averaging volume tends to emphasize the regions in tension above and below the nanoprecipitate along the [0 0 1] crystallographic direction. As already observed in **Figure 2**c, in order to reach the equilibrium $\frac{c}{a}$ ratio, the *c*-axis of the $L1_0$ nanoprecipitate must go into compression during the energy minimization ($\overline{\varepsilon_{001}} \approx -2.6\%$). **Figures 3**b-c show the $\varepsilon_{001}$ strain distribution after energy minimization for the FCC and FCT configurations respectively, computed from the fundamental reflection *g* = 0 0 2, this time taking the perfect FCC lattice as the reference for the strain calculation in both configurations. In contrast to the FCT reference, the RSV is in this case centred around the intensity scattered by the disordered FCC *A1* matrix (**Figure S3**f). In order to obtain a strain distribution consistent with the one obtained in the atomistic configuration, the retrieved $\varepsilon_{001}$ strain component is also averaged over the first five neighboring voxels in all 3 directions of space (**Figures 3**c-d). Hence, the averaging volume is very similar to the one used in atomistic calculations (1.423 nm$^3$ for atomistic simulation *vs* 1.436 nm$^3$ for retrieved strain). Using the FCC reference lattice confirms that the strain distribution is identical for the two configurations, not only for the *A1* phase but also in the $L1_0$ nanoprecipitate ($\overline{\varepsilon_{001}} \approx -2.6\%$). The strain distribution is also very consistent with the atomic strain calculated from the relaxed atomic

positions in the FCC configuration (**Figure 3**a). This confirms that the relaxed state of the nanoprecipitate does not depend on its initial (FCC or FCT) crystal structure: both FCC and FCT precipitate reach an optimum $\frac{c}{a}$ ratio of 0.96 after energy minimization. The discrepancy between the strain states directly obtained from the atomic positions for the FCT and FCC configurations can therefore be explained by the different reference lattices used for the calculations. After relaxation, a tensile strain (**Figures 2**g,i) is obtained for both configurations of nanoprecipitate with the FCT reference, while a compressive strain (**Figures 3**b-c) is observed in both cases with the FCC reference. For the superstructure reflection ($g$ = 0 0 1), one also obtains an average strain value of −2.6% for the nanoprecipitate (**Figure 3**d), while the matrix region is not retrieved accurately, a result in good agreement with the calculations performed with the reference FCT lattice (**Figure 2**i).

From these calculations, one can conclude that the strain distribution retrieved from the diffraction pattern (see *Eqs.* (1) and (2)) is very accurate and consistent with the direct calculations from the atomic positions. Fundamental and superstructure reflections give similar results for the ordered phase, while the strain is retrieved accurately only for the fundamental reflection in the disordered matrix. The equilibrium configuration of the precipitate is independent of its initial crystal structure: the strain distribution after relaxation is identical for the FCT and FCC precipitates, but one needs to be careful when selecting the reference lattice in order to correctly interpret the strain distribution in the nanoprecipitate. The case of semi-coherent and incoherent precipitates is discussed in *supporting information S1*.

### 3.2. Influence of the order parameter on the reconstructed data from a superstructure reflection

In the previous section, we consider fully ordered and defect free precipitates. However, this is not necessarily the case experimentally, in particular at the early stages of the ordering, where a large number of defects such as antiphase boundaries are observed (Warren, 1969). A convenient approach to characterize the degree of ordering is the long range order parameter defined by (Warren, 1969):

$$S = r_\alpha - w_\gamma = r_\gamma - w_\alpha, \qquad (3)$$

where $r_\alpha$ and $r_\gamma$ are the fraction of $\alpha$ and $\gamma$ atomic sites occupied by the right atoms, while $w_\alpha$ and $w_\gamma$ are the fraction of $\alpha$ and $\gamma$ sites occupied by the wrong atoms. From this definition, it comes that $S = 1$ corresponds to a fully ordered crystal with a stoichiometric composition, while $S = 0$ corresponds to a completely random arrangement of atoms. The structure factors $F$ for the superstructure reflections are proportional to $S$ and therefore a $S^2$ parameter can be derived from the measured intensity during experiment.

We consider here that for a completely ordered alloy with ideal stoichiometry, all the $\alpha$ sites are occupied by Cu atoms, while all the $\gamma$ sites are occupied by the Au atoms. The sample composition is the sum of the atom fractions $n_{Cu} + n_{Au} = 1$. The structure factor for a partially ordered alloy can be obtained by summing over all atomic positions in the unit cell. For a fundamental reflection such as the 0 0 2 it is given by:

$$F_{0\,0\,2} = 2[f_{Au}\{r_\gamma + w_\alpha\} + f_{Cu}\{r_\alpha + w_\gamma\}] = 4(f_{Au}n_{Au} + f_{Cu}n_{Cu}), \quad (4)$$

where $f_{Au}$ and $f_{Cu}$ are the Thompson scattering factors.

For a superstructure reflection such as the 0 0 1, the structure factor is given by:

$$F_{0\,0\,1} = 2[f_{Au}\{r_\gamma - w_\alpha\} + f_{Cu}\{-r_\alpha + w_\gamma\}] = 2S(f_{Au} - f_{Cu}). \quad (5)$$

It comes that the integrated intensity of the 0 0 2 fundamental reflection is equal to:

$$I_{002} = 16cV_{002}(f_{Au}n_{Au} + f_{Cu}n_{Cu})^2, \quad (6)$$

where $V_{002}$ is the volume of the region at the Bragg condition and $c$ are the scattering constants. A similar expression can be derived for the 0 0 1 superstructure reflection:

$$I_{001} = 4cV_{001}S^2(f_{Au} - f_{Cu})^2. \quad (7)$$

Assuming that the intensities are measured from a region of reciprocal space with equal volume ($V_{001} = V_{002}$) one can derive the order parameter from the ratio between integrated intensities:

$$\frac{I_{001}}{I_{002}} = \frac{S^2(f_{Au} - f_{Cu})^2}{4(f_{Au}n_{Au} + f_{Cu}n_{Cu})^2}, \quad (8)$$

which gives

$$S = \sqrt{\frac{I_{001}}{I_{002}}} \frac{2(f_{Au}n_{Au} + f_{Cu}n_{Cu})}{(f_{Au} - f_{Cu})} = \frac{A_{001}}{A_{002}} \frac{2(f_{Au}n_{Au} + f_{Cu}n_{Cu})}{(f_{Au} - f_{Cu})} \quad (9)$$

In order to evaluate the influence of the order parameter $S$ on the reconstructed data, the latter was varied in two different ways:

- by modifying the chemical composition of the cell, *i.e.* a fraction of the Au atoms on the $\gamma$ sites is replaced by Cu atoms (hereafter referred as varying composition),

- by keeping constant the stoichiometric composition ($n_{Au} = n_{Cu} = 0.5$), a fraction of the Cu atoms occupies the $\gamma$ sites, while the same fraction of Au atoms occupies the $\alpha$ sites (hereafter referred as fixed composition). Doing so the stoichiometric composition is kept constant but the order parameter $S$ decreases as the fraction of Au atoms (resp. Cu atoms) increases on the $\alpha$ sites (resp. $\gamma$ sites).

**Figure 4** shows the case of a fully ordered cell ($S = 1$) with a stoichiometric composition: all the Au atoms are located on the $\gamma$ sites, while all the Cu atoms are on the $\alpha$ sites. The scattered amplitudes are calculated from equation (1) for a superstructure reflection ($\boldsymbol{g} = 0\,0\,1$) and for a fundamental reflection ($\boldsymbol{g} = 0\,0\,2$) using a small (50x50x66 RSPs) and a large (150x150x200 RSPs) RSV. The calculations are performed on a relaxed configuration. The complex sample densities are then derived from equation -

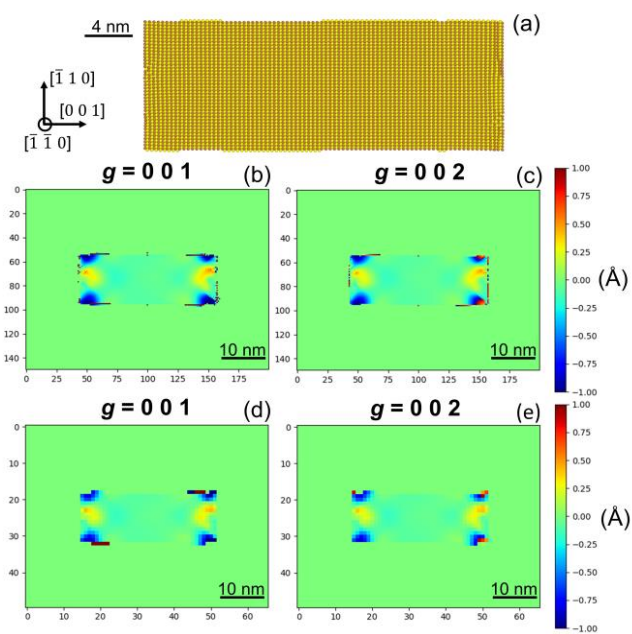

**Figure 4** (a) Atomistic configuration for an ordered FCT $L1_0$ phase (S = 1). $u_{001}$ displacement in the (1 1 0) plane calculated from a superstructure reflection, $g$ = 0 0 1 for a small (0.357 nm / RSV(2.51x2.51x3.64 nm$^{-3}$)) (b) and a large (1.09 nm / RSV(0.827x0.827x1.19 nm$^{-3}$)) (d) voxel size. Same $u_{001}$ displacement calculated from a fundamental reflection, $g$ = 0 0 2 for a small (c) and a large (e) voxel size. The voxels, for wich $\rho < 0.2\ \rho_{max}$, are set to zero.

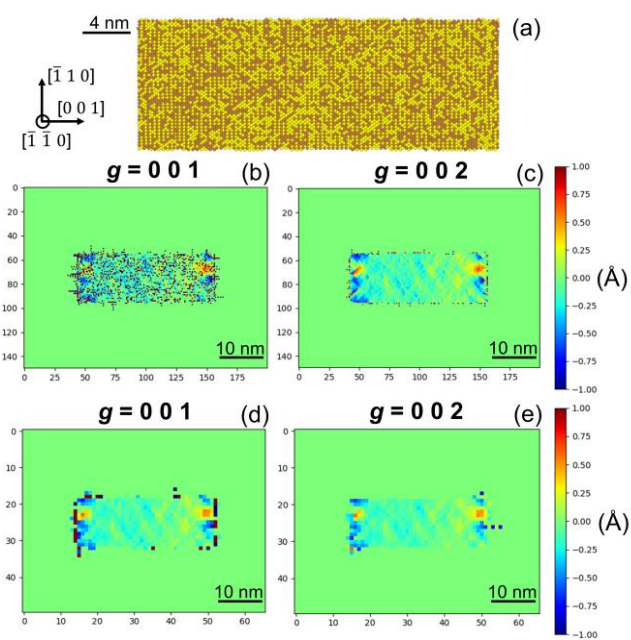

**Figure 5** (a) Atomistic configuration for a partially ordered FCT $L1_0$ phase (S = 0.5). $u_{001}$ displacement in the (1 1 0) plane calculated from a superstructure reflection, $g$ = 0 0 1 for a small (0.357 nm / RSV(2.51x2.51x3.64 nm$^{-3}$)) (b) and a large (1.09 nm nm / RSV(0.827x0.827x1.19 nm$^{-3}$)) (d) voxel size. Same $u_{001}$ displacement calculated from a fundamental reflection, $g$ = 0 0 2 for a small (c) and a large (e) voxel size. The voxels, for wich $r < 0.2\ r_{max}$, are set to zero.

- (2). **Figures 4**b-e show the displacement field component along the *c*-axis, $u_{0\,0\,1}$, calculated for **g** = 0 0 1, (**Figures 4**b,d) and **g** = 0 0 2 (**Figures 4**c,e). The calculations performed on the small (**Figures 4**d,e) and large (**Figures 4**b,c) RSVs both reveal an excellent agreement between the two reflections: the calculation of the Pearson correlation coefficient gives a 99.4% and 99.8% agreement for the small and large RSV, respectively (**Table 1**). This excellent agreement is consistent with the theory as both reflections are sensitive to the same component of the displacement field.

In contrast, **Figure 5** shows the case of a partially ordered cell ($S = 0.5$), where 25% of the $\alpha$ sites are occupied by Au atoms and 25% of the $\gamma$ sites are occupied by Cu atoms. The calculation of $u_{0\,0\,1}$ for a relaxed configuration reveals some discrepancies between the superstructure and the fundamental reflections. This is especially true for the large RSV, where the voxel size (0.357 nm) is just slightly larger than the first neighbour distance (0.276 nm, **Figures 5**b,c). Since the Au (resp. Cu) atoms are placed randomly on the $\alpha$ (resp. $\gamma$) sites, the local ordering varies significantly depending on the location in the crystal. In some regions, the signal from the superstructure reflection is very weak, resulting in an extremely low electron density. In these regions, the $u_{0\,0\,1}$ is not retrieved accurately as shown in **Figure 5**b. For the small RSV (voxel size 1.09 nm) on the other hand (**Figures 5**d,e), the agreement between the fundamental and the superstructure reflection remains excellent. Because each voxel contains on average 90 atoms (only 3 atoms per voxel for the large), the local ordering is less dependent on the atomic position and the retrieved electron density is reasonably high and homogeneous everywhere in the crystal. As a consequence, $u_{0\,0\,1}$ is still retrieved accurately for **g** = 0 0 1 *(***Figure 5**d). The calculation of the Pearson correlation confirms the visual interpretation with calculated correlations of 91.5% and 59.6% for the small and large windows, respectively (see also **Figure 9** & **Table S14**). In order to quantify the accuracy of the strain and displacement retrieved from the superstructure reflections, we calculated the evolution of the ratio of the integrated electron density retrieved in the real space ($\rho_{int\_001} / \rho_{int\_002}$), and the ratios of the integrated scattered amplitude ($A_{int\_001} / A_{int\_002}$) and intensities ($I_{int\_001} / I_{int\_002}$) in the reciprocal space as a function of the order parameter S. These values are compared with the theoretical ratio ($I_{th\_001} / I_{th\_002}$) and ($A_{th\_001} / A_{th\_002}$) calculated from equation (8) (see **Tables S3-S8**). As a reminder, the amplitudes and intensities are typically integrated over 25x25x33 and/or 50x50x66 RSPs centred around the maximum of intensity of the calculated Bragg reflection. For the integrated electron densities, three thresholds are considered (25%, 32.5% and 40% of the maximum electron density).

**Figure 6** shows the evolution of the intensities integrated over 25x25x33 RSPs for a varying and a fixed composition, respectively. The results for the large RSV, where the integration can be performed over larger RSVs (up to 150x150x200), are presented in *supporting information S4-S6*. The calculations for the varying composition reveal an excellent agreement with the theory for both relaxed and unrelaxed configurations (**Figure 6**a & **Table S3**). One can also note that ($I_{int\_001} / I_{int\_002}$) is mostly independent of the selected RSV for integration. The agreement remains very good for a fixed composition; however, one can notice that larger deviations are observed for a low order parameter ($S \leq 0.2$), where the ratio of integrated intensities is significantly larger than the theoretical value (**Figure 6**b & **Table S4**).

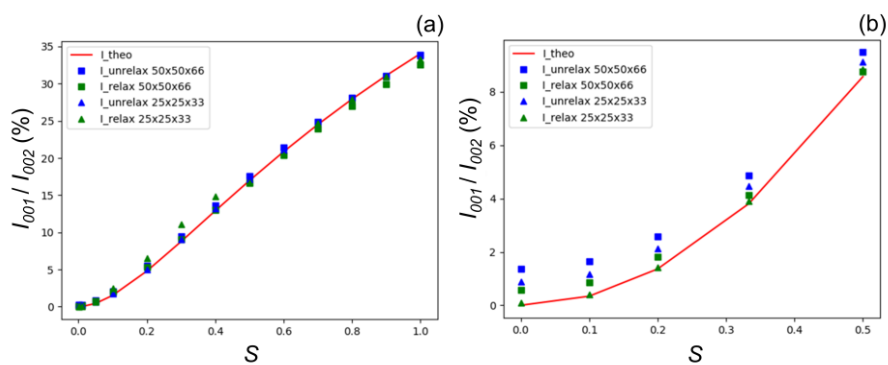

**Figure 6** Evolution of $(I_{int\_001} / I_{int\_002})$ and $(I_{th\_001} / I_{th\_002})$ for relaxed and unrelaxed configurations as a function of the order parameter *S* for a varying composition (a) and a fixed composition (b). In both cases, the calculations are carried out on the small RSV (50x50x66 RSPs) and the integrations are performed over 25x25x33 and 50x50x66 RSPs.

Interestingly, these deviations are more pronounced for the unrelaxed configurations. In addition, $I_{int\_001} / I_{int\_002}$ shows a larger dependence to the size of the RSV, the smaller RSV showing the best agreement with the theory. We assume that the larger deviations for the unrelaxed configurations are most likely related to artefacts caused by the finite size of the simulation box.

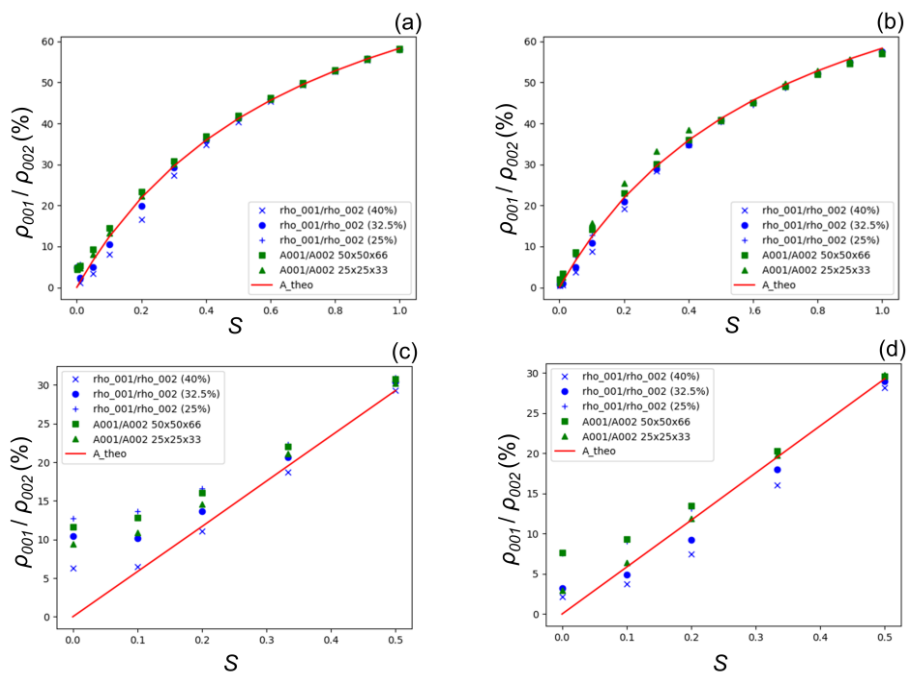

**Figure 7** Evolution of $(A_{int\_001} / A_{int\_002})$, $(A_{th\_001} / A_{th\_002})$ and $(\rho_{int\_001} / \rho_{int\_002})$ as a function of the order parameter *S* for a varying atomic composition before (a) and after (b) relaxation. Same ratios as a function of *S* for the fixed composition before (c) and after (d) relaxation. In both cases, the calculations are carried out on the small RSV (50x50x66 RSPs) and the integrations are performed over 25x25x33 and 50x50x66 RSPs.

Similar observations can be drawn from the calculation of ($A_{int\_001}$ / $A_{int\_002}$), which enables a direct comparison with ($\rho_{int\_001}$ / $\rho_{int\_002}$) (**Figure 7**). As shown in **Figures 7**a-b and **Tables S5-S8**, the ($\rho_{int\_001}$ / $\rho_{int\_002}$) and ($A_{int\_001}$ / $A_{int\_002}$) ratios are almost in perfect agreement for the varying composition. This can be understood from mathematical considerations since the electron density is the modulus of the Fourier transform of the scattered amplitude (equation (2)). Logically, they are also in very good agreement with the theoretical ratio ($A_{th\_001}$ / $A_{th\_002}$); the small deviations already reported for the integrated intensities for low $S$ values ($S \leq 0.15$) are here well visible, in particular for the unrelaxed configurations (**Figure 7**a). We also note that the results are extremely robust since they are mostly independent of the integration volume for the scattered amplitude and of the threshold used for the integration of the electron density.

For the fixed composition, ($\rho_{int\_001}$ / $\rho_{int\_002}$) and ($A_{int\_001}$ / $A_{int\_002}$) are also in very good agreement (**Figures 7**c-d & **Table S6**). However, the ($\rho_{int\_001}$ / $\rho_{int\_002}$) ratio shows a larger dependence to the integration threshold than for the varying composition: the larger is the threshold, the smaller is the ($\rho_{int\_001}$ / $\rho_{int\_002}$) ratio. The smaller is the RSV, the smaller is the ($A_{int\_001}$ / $A_{int\_002}$) ratio. The trends observed for ($\rho_{int\_001}$ / $\rho_{int\_002}$) are similar to the ones observed for the integrated intensities and amplitudes; the largest deviations are observed at low $S$, the ratios calculated for the unrelaxed configurations are significantly larger than the theory ($A_{th\_001}$ / $A_{th\_002}$), while a better agreement is obtained for the relaxed configurations. In addition, a much better agreement is obtained for the small RSV / large voxel size (**Figure 7**c-d & **Table S6**) than for the large RSV / small voxel size (**Figures S5**c,d & **Table S8**).

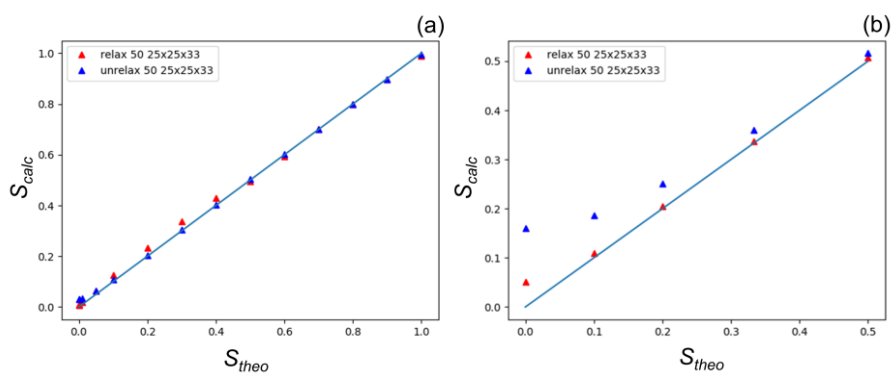

**Figure 8** Evolution of the order parameter calculated from the ratio of the integrated intensities ($S_{calc}$, equation (9)) as a function of the theoretical order parameter ($S_{theo}$, equation (3)) for the varying (a) and fixed (b) compositions. In both cases, the calculation is performed on the small RSV (50x50x66 RSPs) and the integrations are performed over 25x25x33 RSPs.

Another way to interpret these results consists in using equation (9), in order to calculate the order parameter ($S_{calc}$) and compare it with its theoretical value ($S_{theo}$) obtained from equation (3). As expected, $S_{calc}$ is in almost perfect agreement with the theoretical value for a varying composition, despite a small deviation for the smaller value of $S$ (**Figure 8**a & **Tables S6**a,c). For a fixed composition, the agreement between $S_{calc}$ and $S_{theo}$ is significantly better for the relaxed configurations than for the unrelaxed, $S_{calc}$ being overestimated for the latter. In addition, the largest deviations are observed in both cases for small $S$ values (**Figure 8**b & **Tables S9,S11**).

Finally, **Figure 9** and **Tables S10,S12** show the evolution of the Pearson correlation $r$ between the $\varepsilon_{001}$ strain calculated from the fundamental and superstructure reflections as a function of the order parameter. In good agreement with our previous observations, this parameter largely depends on the real space voxel size. Indeed, for a large voxel size (small RSV), $r$ remains high even for a very low order parameter, while $r$ drops rapidly with $S$ for a small voxel size (large RSV). This trend is well visible for the varying composition for both relaxed and unrelaxed configurations (**Figure 9**a & **Table S13**). For the large voxel size, $r$ remains above 90% in the range $0.1 < S < 1$, while for a small voxel size it drops from nearly 100% ($S = 1$) to around 30% for $S = 0.1$ (34% and 28% for the relaxed and unrelaxed configurations, respectively). Interestingly for the small voxel size in the range $0.0002 < S < 0.1$, $r$ remains very high and above 90% for the unrelaxed configuration, while a sharp drop is observed for the relaxed configurations (from $r = 92.3\%$ for S = 0.1 to $r = 48.9\%$ for S = 0.0002). The origin of this discrepancy remains unclear and could be caused by artificial ordering in the unrelaxed configuration due to the small size of the simulation box. Similar trends are observed for the fixed composition, although $r$ shows less dependence to the order parameter, even for the small voxel size, where $r$ drops by only 20%, when decreasing $S$ from 0.5 to 0 (**Figure 9**b & **Table S14**). Similarly, to the varying composition, for a large voxel size and for the unrelaxed configurations, $r$ remains very high (~ 90%) and mostly independent on the order parameter. In contrast, for the relaxed configurations $r$ drops significantly (by ~ 20%) for low $S$ values.

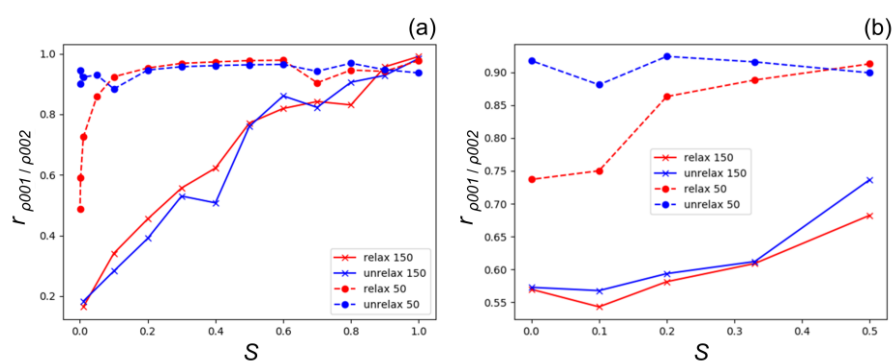

**Figure 9** Pearson correlation between the $\varepsilon_{001}$ retrieved from the superstructure ($\mathbf{g}$ = 0 0 1) and the fundamental ($\mathbf{g}$ = 0 0 2) reflections as a function of the order parameter $S$ for the varying (a) and fixed (b) compositions for the relaxed and unrelaxed configurations and different integrated RSVs.

Similar trends are observed for the fixed composition, although $r$ shows less dependence to the order parameter, even for the small voxel size, where $r$ drops by only 20%, when decreasing $S$ from 0.5 to 0 (**Figure 9**b & **Table S14**). Similarly, to the varying composition, for a large voxel size and for the unrelaxed configurations, $r$ remains very high (~ 90%) and mostly independent on the order parameter. In contrast, for the relaxed configurations $r$ drops significantly (by ~ 20%) for low $S$ values.

In summary, one can conclude that the larger is the voxel size, the better is the accuracy of the retrieved strain. For a voxel size larger than 1 nm, the strain retrieved for low order parameter is still in excellent agreement with the calculations from the relaxed atomic positions, even $S$ drops to low values. This is especially true when the decrease of the order parameter is induced by a variation of the composition. In addition, a better agreement is generally obtained for relaxed configurations compared to unrelaxed configurations. These two results are encouraging and promising in the optic of an experimental validation of the method:

- The resolution achieved experimentally is approximately one order of magnitude larger than the voxel size range used in these simulations. The accuracy of the retrieved strain should greatly benefit from this larger averaging volume.
- The relaxed configurations, where large and inhomogeneous strains are observed in the precipitate and in the matrix, are more representative of the experimental samples.

For these two reasons the technique appears to be suitable not only for perfect/ideal samples but also for realistic defective/strained nanoprecipitates with a low order parameter which are found in many systems.

### 3.3. Imaging of an assembly of nanoprecipitates

In the first two subsections, we have seen that the real space displacement and strain fields can be retrieved very accurately by performing an inverse Fourier transform on the complex scattered amplitude (equation (2)). A perfect agreement is obtained with the atomic strains calculated directly from the relaxed atomic positions (**Figure 2** & **Figure 3**). Moreover, the influence of the order parameter on the reconstructed data was evaluated quantitively (subsection 3.2). It was established that the accuracy of the retrieved strain from the superstructure reflections benefits from large averaging volumes (voxel size) and is also usually better for relaxed configurations; two important results if one aims at using the method on experimental samples. However, if the complex sample density can be easily derived from the scattered amplitude in simulations, using equation (2), this is not the case experimentally, where a scattered intensity is measured on the far-field detector. The present section aims at establishing the ability of phase retrieval algorithms to recover the complex sample density from high resolution diffraction patterns measured in the vicinity of a superstructure reflection. To do so, we considered several crystals with a varying number $N$ of $L1_0$ nanoprecipitates ($N$ varying from 1 to 48) coherently embedded in the $A1$ matrix phase. As indicated in the methods section and as shown in **Table 1**, the crystals size range from 11x11x11 nm$^3$ to 66x66x66 nm$^3$ and the sampling conditions in the reciprocal space are adjusted accordingly to fulfil the oversampling conditions. **Figure 10**a shows a 66x66x66 nm$^3$ crystal containing an assembly of 24 FCT $L1_0$ nanoprecipitates randomly distributed in the matrix solid solution. The radii of the precipitates vary between 1.5 and 2.8 nm. Note that some nanoprecipitates are rotated by few degrees around their $c$-axis and are therfore not fully coherent with the matrix. The atoms of the $A1$ matrix phase are removed in order to allow better visualization of the nanoprecipitates. **Figure 10**b shows the evolution of the $\varepsilon_{001}$ atomic strain component after energy minimization. One can notice the large and local strain variations at the matrix/nanoprecipitates interfaces. The scattered intensity from this relaxed configuration (including the $A_1$ matrix atoms) was calculated using equation (1) in the vicinity of the 0 0 1 superstructure reflection ($g = 0\ 0\ 1$) and used as input to evaluate the ability of BCDI to reconstruct the complex image of an assembly of precipitates.

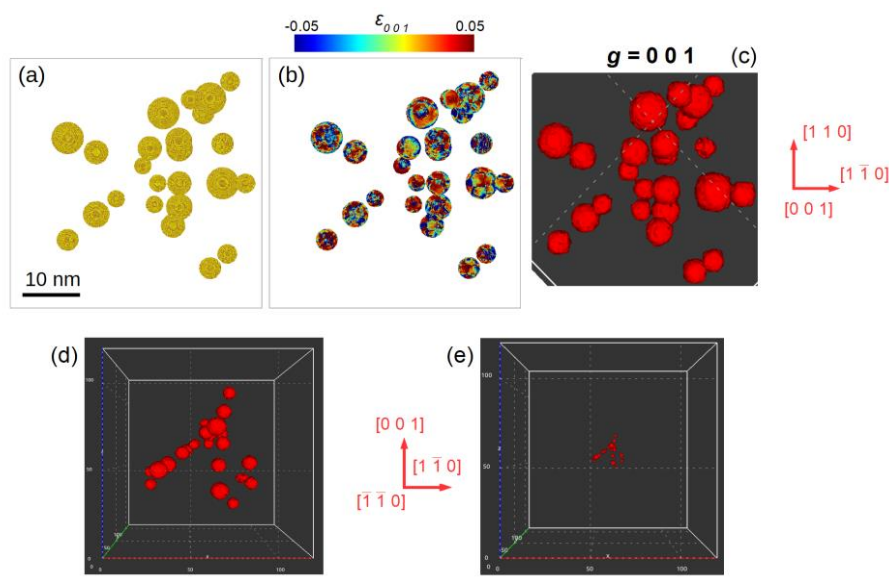

**Figure 10** (a) Simulation cell containing an assembly of 24 $L1_0$ nanoprecipitates (r varying from 1.5 to 2.8 nm). The atoms from the $A_1$ matrix phase are not shown in order to facilitate the visualization. (b) Evolution of $\varepsilon_{001}$ after relaxation. (c-d) Isosurface of the reconstructed Bragg electron density in the vicinity of the 0 0 1 superstructure reflection (RSV: 1.25x1.25x1.25 nm$^{-3}$ / real space voxel size: 0.731 nm). The threshold for the isosurface is set to 42% of the maximum of the Bragg electron density. (c) is a zoom of (d). (e) Reconstructed Bragg electron density for the same reflection ($g = 0\ 0\ 1$) and a smaller RSV / larger voxel size (RSV:0.339x0.339x0.339 nm$^{-3}$ / real space voxel size: 2.91 nm). The threshold for the isosurface is set to 45% of the maximum of the Bragg electron density

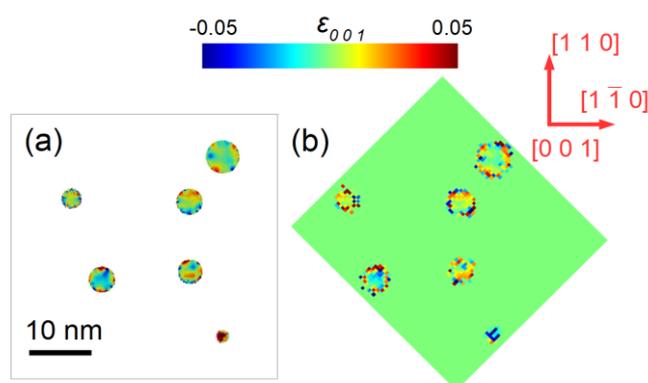

**Figure 11** (a) $\varepsilon_{001}$ atomic strain component displayed on a 1 nm thick (4 atomic layers) (0 0 1) slice taken at the centre of the relaxed configuration. A 0.365 nm cut-off radius is selected for the averaging, so that the averaging volume matches the voxel size of the phase retrieval data. The atoms from the $A1$ matrix are discarded to ease the visualization of the data (b) (0 0 1) slice of the reconstructed $\varepsilon_{001}$ strain component retrieved from the 0 0 1 superstructure reflection (RSV: 1.25x1.25x1.25 nm$^{-3}$ / real space voxel size: 0.731 nm).

A close-up view of the intensity scattered by a varying number in the vicinity of ($\boldsymbol{g}$ = 0 0 1) is shown in *supporting information S9*. The result of the phase retrieval using the procedure detailed in the methods section is show in **Figures 10**c-e. **Figures 10**c-d show the retrieved Bragg electron density for a 0.731 nm voxel size: a perfect agreement is found with the atomistic configuration indicating that the complex electron density was succesfully retrieved.

**Table 1** Influence of the number of precipitates, oversampling conditions, and voxel size on the success rate of the phase retrieval. The lines in black refer to the conditions with a high oversampling ratio (σ >80) while the ones in red indicate conditions with a low oversampling ratio (σ < 32)

|  | N precipitates | RSV (nm$^{-3}$) | Crystal size (nm$^3$) | Oversampling | Radius precipitates (nm) | Voxel size (nm) | Success rate (%) |
|---|---|---|---|---|---|---|---|
| Unrelaxed | 1 | 3.354x3.354x3.630 | 11x11x11 | 82 | 1.755 | 0.291 | 96 |
|  | 1 | 2.264x2.264x2.264 | 11x11x11 | 256 | 1.755 | 0.442 | 100 |
|  | 1 | 2.684x2.684x2.904 | 33x33x33 | 5.5 | 5.265 | 0.363 | 90 |
|  | 1 | 1.004x1.004x1.087 | 44x44x44 | 17.8 | 7.5 | 0.971 | 84 |
|  | 1 | 0.508x0.508x0.508 | 44x44x44 | 140.5 | 7.5 | 1.95 | 100 |
|  | 2 | 2.236x2.236x2.420 | 22x22x22 | 31 | 2.5 | 0.436 | 99 |
|  | 2 | 1.359x1.359x1.359 | 22x22x22 | 145 | 2.5 | 0.736 | 89 |
|  | 3 | 1.359x1.359x1.359 | 22x22x22 | 145 | 2.5 | 0.736 | 85 |
|  | 4 | 2.034x2.034x2.034 | 22x22x22 | 18.7 | 2.5 | 0.492 | 84 |
|  | 5 | 2.034x2.034x2.034 | 22x22x22 | 18.7 | 2.5 | 0.492 | 87 |
|  | 8 | 2.034x2.034x2.034 | 22x22x22 | 18.7 | 2.5 | 0.492 | 88 |
|  | 24 | 1.004x1.004x1.087 | 66x66x66 | 5.2 | 1.5-2.8 | 0.971 | 84 |
|  | 48 | 0.872x0.872x0.872 | 66x66x66 | 8.4 | 1.5-2.8 | 1.15 | 82 |
|  | 48 | 0.339x0.339x0.339 | 66x66x66 | **138.7** | 1.5-2.8 | 2.95 | **89** |
| Relaxed | 1 | 3.354x3.354x3.630 | 11x11x11 | 82 | 1.755 | 0.291 | 94 |
|  | 1 | 2.264x2.264x2.264 | 11x11x11 | 255 | 1.755 | 0.442 | 85 |
|  | 2 | 1.359x1.359x1.359 | 22x22x22 | 144 | 2.5 | 0.736 | 100 |
|  | 3 | 1.359x1.359x1.359 | 22x22x22 | 144 | 2.5 | 0.736 | 87 |
|  | 4 | 2.034x2.034x2.034 | 22x22x22 | 18.4 | 2.5 | 0.492 | 96 |
|  | 5 | 2.034x2.034x2.034 | 22x22x22 | 18.4 | 2.5 | 0.492 | 99 |
|  | 8 | 2.034x2.034x2.034 | 22x22x22 | 18.4 | 2.5 | 0.492 | 92 |
|  | 24 | 1.004x1.004x1.087 | 66x66x66 | 5.2 | 1.5-2.8 | 0.971 | 87 |
|  | 48 | 0.872x0.872x0.872 | 66x66x66 | 8.4 | 1.5-2.8 | 1.15 | 75 |

| | | | | | | |
|---|---|---|---|---|---|---|
| *48* | *0.339x0.339x0.339* | *66x66x66* | **138.7** | *1.5-2.8* | *2.95* | *83* |

As illustrated in **Figure 11**, which shows a (0 0 1) slice at the centre of the same configuration, the strain magnitude and distribution in the precipitates (**Figure 11**b) is also very consistent with the atomic strain computed from the relaxed atomic positions (**Figure 11**a). The same procedure was repeated on a smaller RSV (0.339x0.339x0.339 nm$^{-3}$) corresponding to a real space voxel size of 2.95 nm. Such voxel size is of the same order of magnitude, although slightly smaller, as the resolution accessible experimentally (Labat *et al.*, 2015; Cherukara *et al.*, 2018). Using this large voxel size does not have a negative impact on the success rate of the phase retrieval procedure (**Table 1**).

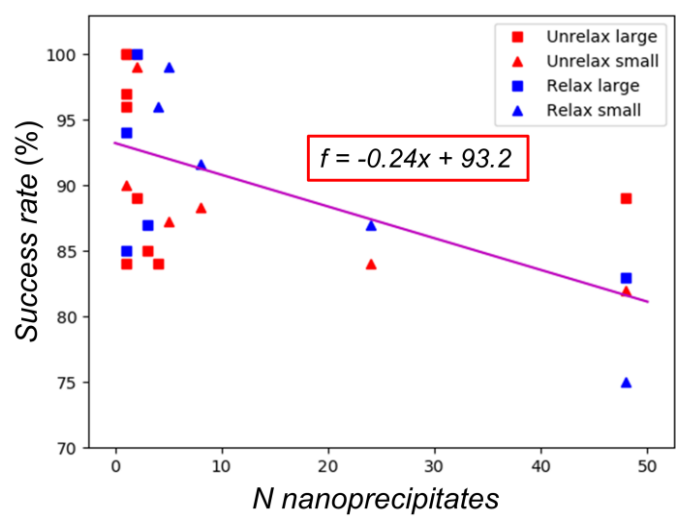

**Figure 12** Evolution of the phase retrieval success rate as a function of the number of nanoprecipitates. The squares / triangles correspond to reconstructions performed with a low ($\sigma < 32$) / high ($\sigma > 80$) oversampling respectively.

**Figure 12** illustrates the success rate of the phasing as a function of the number of nanoprecipitates in the simulation cell. Interestingly, the success rate remains high (> 75%) even for configurations containing a large number of precipitates (**Figure 12** & **Table 1**). However, it is also clear from **Table S15** that a successful phasing can only be achieved if the threshold for shrink-wrap is set properly. As shown from the negative slope of the linear fit, the success rate is not completely independent of the number of nanoprecipitates and is in fact a decreasing function of the latter. However, the current work features only a limited number of data points, especially for large *N* (as illustrated in **Figure 12** where 26 simulations/data points were computed in total but only 6 for *N* > 10). To conclude on the dependence of the success rate on the number of precipitates would require performing a more systematic study with at least one order of magnitude more data points.

Finally, **Table 1** allows to evidence that the success rate is mostly independent on the oversampling ratio: a similar success rate is obtained for $\sigma < 32$ (90 %) and for $\sigma > 80$ (90.9 %). Additionally, a similar success rate is obtained for the relaxed (89.8%) and for the unrelaxed (90.9%) configurations. The large strains induced during the relaxation are therefore not detrimental for the phasing.

In conclusion of this section we have shown that:

- Superstructure reflections can be used to perform BCDI.

- The success rate of phasing remains high even for a large number of precipitates.

These two results are important since to the best of our knowledge, BCDI using superstructure reflections has never been demonstrated experimentally, and more importantly BCDI is typically employed on isolated single objects. In the present work, we demonstrated that it can also be successfully employed to image an assembly of nanoobjects which could open new avenues for the technique.

## 4. Discussion

Coherent precipitation and more generally ordered phases are beneficial not only for the HT mechanical properties of CCAs but can also enhance the magnetic or catalytic properties of alloyed nanoparticles. In bulk CCAs, not only the volume fraction, size and distribution of the nanoprecipitates but also their shape have a large influence on the mechanical properties of these alloys. Because the misfit strain strongly influences the HT properties, the detailed knowledge of the precipitate shape and 3D elastic strain and of their dynamical evolution during ageing is of crucial importance for the development of new alloys. The good spatial resolution and extremely high strain sensitivity of BCDI make it the only technique suitable to image the 3D field of nanoprecipitates in the size range where they are still in perfect coherency with the matrix. More importantly the technique is particularly adapted for *in situ* and *operando* experiments (Dupraz *et al.*, 2017; Kim *et al.*, 2018) foreseeing the possibility to image the evolution of the shape and strain fields in real time during ageing at HT. Such knowledge would help understanding how to design stable microstructures that retain excellent mechanical properties at HT. Going one step further, one could perform *in situ* deformation of bulk or thin film specimens. Such experiments would allow investigating the interaction of dislocation with the elastic strain field of precipitates and better understand how the latter affect the mobility of dislocations.

BCDI would be also particularly relevant for the study of ordered particles where both fundamental and superstructure reflections can be used to gain further in the 3D strain distribution of isolated nanoparticles. Moreover, the possibility to use the technique *in situ* and *operando* makes it a potentially powerful tool to monitor the nucleation of the ordered phase and the kinetics of ordering during the coarsening of the ordered phase. In such systems, superstructure and fundamental reflections can provide very complementary information. The former can be used to precisely locate and investigate potential defect structure in the ordered phase such as anti-phase boundaries that are quite common in the early stage of ordering. Indeed, only superstructure reflections are sensitive to these phase defects (Warren, 1969).The fundamental reflections on the other hand would provide information on the extent of the strain field of the precipitates in the disordered matrix. In catalytic nanoparticles such as $PtCu_3$ the technique could contribute to improve our knowledge on the relationship between strain, crystal defects and catalytic activity which is essential in order to design better catalysts.

Most of the aforementioned experiments in bulk CCAs or in ordered nanoparticles require spatial resolution of a few nanometres. Such values correspond roughly to the current limits of the technique. In addition, the imaging of the local strains in $L1_0$ spherical nanoprecipitates presented in this study requires to achieve a spatial resolution of the order of 1 nm. Currently out-of-reach for state of the art BCDI which can only achieve a spatial resolution at best 5 times higher, next generation synchrotron sources will look to bridge the gap. However, the results presented in this work can be generalized to larger precipitates sizes that are routinely encountered in many systems (Ni-based superalloys, Fe-based alloys, HEAs, …). In such systems, coherency is generally retained for a precipitate size up to 500 nm. Their typical size of few hundreds of nanometres is already perfectly suitable for the current capabilities of BCDI. Additionally, the technique will benefit from the upgrades of the 3$^{rd}$ generation synchrotron that was initiated this year with the ESRF-EBS (Extremely Brilliant Source) upgrade. Taking advantage of the increase of the brilliance and coherence of the X-ray beams, one can expect to gain at least a factor of three in resolution which would open new experimental avenues. In addition, we have seen that the success rate of the phase retrieval is independent of the voxel size: precipitates which contains very few voxels can still be reconstructed accurately.

The limited experimental coherent flux also questions the ability to reconstruct displacement and strain fields from superstructure reflections. The scattered intensity from superstructure reflections, is indeed much weaker than the scattering from fundamental reflections. Because of that, BCDI on superstructure will greatly benefit from the EBS upgrade which will allow to investigate new systems and tackle new fundamental questions. Still, some samples are more suitable to the technique than others. For a given order parameter, the superstructure reflections are much more intense for the $L1_0$ and $B2$ ordered phases than for the $L1_2$. In addition, alloys containing elements with large difference in their atomic number $Z$ (which implies a large difference in the Thompson scattering factor) are more suitable for the technique. For instance, at 8keV a strong scattering can be expected from a fully ordered ($S = 1$) PtNi $L1_0$ phase:

$$\frac{I_{001}}{I_{002}} = \frac{S^2(f_{Pt} - f_{Ni})^2}{4(f_{Pt}n_{Pt} + f_{Ni}n_{Ni})^2} = 0.23$$

On the other, the scattering from the ordered Ni$_3$Fe $L1_2$ phase is extremely weak and would be challenging to measure experimentally even for $S = 1$:

$$\frac{I_{001}}{I_{002}} = \frac{S^2(f_{Fe} - f_{Ni})^2}{16(f_{Fe}n_{Fe} + f_{Ni}n_{Ni})^2} = 0.00049$$

Finally, similarly to other diffraction techniques, BCDI is a non-destructive technique which takes advantage of the large penetration of X-rays in the matter to image the strain and displacement fields without requiring extensive sample preparation like TEM or APT. In this respect, it does not suffer from the free surfaces boundary conditions that can favour deformation mechanisms not representative of the bulk samples in TEM lamella. That being said, an extensive sample preparation is still required if one wants to reconstruct the strain field of bulk samples with a grain size larger than 1 mm using fundamental reflections (Hofmann *et al.*, 2020). We have seen in this work that the use of superstructure reflections could relax this constraint since the finite size of the nanoprecipitates can help the convergence of the phase retrieval algorithms. However, for single crystal and coarse-grained samples, the longitudinal coherence length of the X-ray beam will most likely constitute the upper limit for the maximum thickness of the sample accessible experimentally. If the latter exceeds the coherence length of the sample, partial coherence effects will probably be difficult to handle by phase retrieval algorithm. Quantifying this partial coherence effects and their influence on the success rate of the phase retrieval algorithms could be the object of a future work. Despite these limitations, a sample thickness comparable to the longitudinal coherence length ($\approx 0.5$ µm at 8 keV on the ID01 beamline of the ESRF for instance) is still one order magnitude larger than the typical thickness of TEM lamella, implying that the deformation mechanisms should be much more representative of the bulk samples.

## 5. Conclusion

We carried out a detailed numerical analysis aimed at evaluating the relevance of BCDI to image coherent precipitates and more generally ordered phase. We have first shown that a fully accurate strain distribution can be retrieved from both fundamental (in ordered and disordered phases) and superstructure (only in ordered phases) reflections. We have also demonstrated that the strain distribution retrieved from superstructure reflections is still very precise for partially ordered phases with large and inhomogeneous strains, in particular for voxel sizes (averaging volume) comparable to the spatial resolution experimentally achievable. In the last section, we have also seen that superstructure reflections can be used to perform BCDI on samples containing a large number (up to 50) of nanoprecipitates. The success rate of phase retrieval was mostly independent of the number of nanoprecipitates for the samples considered in this study. This could open new avenues for the technique as BCDI is typically used on single isolated objects.

Finally, the technique will definitely benefit from the multiple upgrades currently being carried out or planned at several third-generation sources. Next-generation sources will provide improved brilliance and thus coherent flux which makes it very feasible to transpose our simulation results to the experiment. These new capabilities could open the door to BCDI as a microscopy tool to study complex real-word materials.

**Acknowledgments**

M.D and M.-I.R. acknowledge financial support from ANR Charline (ANR-16-CE07-0028-01) and ANR Tremplin ERC (ANR-18-ERC1-0010-01). This project has received funding from the European Research Council (ERC) under the European Union's Horizon 2020 research and innovation programme (grant agreement No. 818823).

**References**

Abinandanan, T. A. & Johnson, W. C. (1993). *Acta Metallurgica et Materialia*. **41**, 17–25.

Andersen, S. J., Zandbergen, H. W., Jansen, J., TrÆholt, C., Tundal, U. & Reiso, O. (1998). *Acta Materialia*. **46**, 3283–3298.

Ardell, A. J. & Nicholson, R. B. (1966). *Acta Metallurgica*. **14**, 1295–1309.

Bhat, M. S., Garrison, W. M. & Zackay, V. F. (1979). *Materials Science and Engineering*. **41**, 1–15.

Biswas, A., Siegel, D. J., Wolverton, C. & Seidman, D. N. (2011). *Acta Materialia*. **59**, 6187–6204.

Booth-Morrison, C., Dunand, D. C. & Seidman, D. N. (2011). *Acta Materialia*. **59**, 7029–7042.

Cadeville, M. C., Dahmani, C. E. & Kern, F. (1986). *Journal of Magnetism and Magnetic Materials*. **54–57**, 1055–1056.

Cahn, J. W. & Lärché, F. (1982). *Acta Metallurgica*. **30**, 51–56.

Charpagne, M.-A., Vennéguès, P., Billot, T., Franchet, J.-M. & Bozzolo, N. (2016). *Journal of Microscopy*. **263**, 106–112.

Cherukara, M. J., Cha, W. & Harder, R. J. (2018). *Appl. Phys. Lett.* **113**, 203101.

Clouet, E., Laé, L., Épicier, T., Lefebvre, W., Nastar, M. & Deschamps, A. (2006). *Nature Mater*. **5**, 482–488.

Daw, M. S. & Baskes, M. I. (1983). *Phys. Rev. Lett.* **50**, 1285–1288.

Daw, M. S. & Baskes, M. I. (1984). *Phys. Rev. B*. **29**, 6443–6453.

Ding, Q., Li, S., Chen, L.-Q., Han, X., Zhang, Z., Yu, Q. & Li, J. (2018). *Acta Materialia*. **154**, 137–146.

Doi, M., Miyazaki, T. & Wakatsuki, T. (1985). *Materials Science and Engineering*. **74**, 139–145.

Dupraz, M., Beutier, G., Cornelius, T. W., Parry, G., Ren, Z., Labat, S., Richard, M.-I., Chahine, G. A., Kovalenko, O., De Boissieu, M., Rabkin, E., Verdier, M. & Thomas, O. (2017). *Nano Lett.* **17**, 6696–6701.

Favre-Nicolin, V., Coraux, J., Richard, M.-I. & Renevier, H. (2011). *J Appl Cryst*. **44**, 635–640.

Fienup, J. R. (1982). *Appl. Opt., AO*. **21**, 2758–2769.

Foiles, S. M., Baskes, M. I. & Daw, M. S. (1986). *Phys. Rev. B*. **33**, 7983–7991.

Garcia-Gonzalez, M., Van Petegem, S., Baluc, N., Hocine, S., Dupraz, M., Lalire, F. & Van Swygenhoven, H. (2019). *Scripta Materialia*. **170**, 129–133.

Garcia-Gonzalez, M., Van Petegem, S., Baluc, N., Dupraz, M., Honkimaki, V., Lalire, F. & Van Swygenhoven, H. (2019). *Acta Materialia*. **191**, 186–197.

Gerchberg, R. W. (1972). *Optik*. **35**, 237.

Giamei, A. F. & Anton, D. L. (1985). *MTA*. **16**, 1997–2005.

Gladman, T. (1999). *Materials Science and Technology*. **15**, 30–36.

Glatzel, U. & Feller-Kniepmeier, M. (1989). *Scripta Metallurgica*. **23**, 1839–1844.

Goerler, J. V., Lopez-Galilea, I., Mujica Roncery, L., Shchyglo, O., Theisen, W. & Steinbach, I. (2017). *Acta Materialia*. **124**, 151–158.

Hodnik, N., Bele, M., Rečnik, A., Logar, N. Z., Gaberšček, M. & Hočevar, S. (2012). *Energy Procedia*. **29**, 208–215.

Hofmann, F., Phillips, N. W., Das, S., Karamched, P., Hughes, G. M., Douglas, J. O., Cha, W. & Liu, W. (2020). *Phys. Rev. Materials*. **4**, 013801.

Jensen, J. K., Welk, B. A., Williams, R. E. A., Sosa, J. M., Huber, D. E., Senkov, O. N., Viswanathan, G. B. & Fraser, H. L. (2016). *Scripta Materialia*. **121**, 1–4.

Jiang, S., Wang, H., Wu, Y., Liu, X., Chen, H., Yao, M., Gault, B., Ponge, D., Raabe, D., Hirata, A., Chen, M., Wang, Y. & Lu, Z. (2017). *Nature*. **544**, 460–464.

Jiao, Z. B., Luan, J. H., Miller, M. K., Yu, C. Y. & Liu, C. T. (2015). *Acta Materialia*. **84**, 283–291.

Jiao, Z. B., Luan, J. H., Zhang, Z. W., Miller, M. K. & Liu, C. T. (2014). *Scripta Materialia*. **87**, 45–48.

Johnson, W. C. & Voorhees, P. W. (1992). Elastically-Induced Precipitate Shape Transitions in Coherent Solids Trans Tech Publications Ltd.

Kaufman, M. J., Voorhees, P. W., Johnson, W. C. & Biancaniello, F. S. (1989). *MTA*. **20**, 2171–2175.

Kim, D., Chung, M., Carnis, J., Kim, S., Yun, K., Kang, J., Cha, W., Cherukara, M. J., Maxey, E., Harder, R., Sasikumar, K., Sankaranarayanan, S. K. R. S., Zozulya, A., Sprung, M., Riu, D. & Kim, H. (2018). *Nat Commun*. **9**, 1–7.

Kitakami, O., Shimada, Y., Oikawa, K., Daimon, H. & Fukamichi, K. (2001). *Appl. Phys. Lett.* **78**, 1104–1106.

Klemmer, T. J., Liu, C., Shukla, N., Wu, X. W., Weller, D., Tanase, M., Laughlin, D. E. & Soffa, W. A. (2003). *Journal of Magnetism and Magnetic Materials*. **266**, 79–87.

Klemmer, T. J., Shukla, N., Liu, C., Wu, X. W., Svedberg, E. B., Mryasov, O., Chantrell, R. W., Weller, D., Tanase, M. & Laughlin, D. E. (2002). *Appl. Phys. Lett.* **81**, 2220–2222.

Klobes, B., Korff, B., Balarisi, O., Eich, P., Haaks, M., Kohlbach, I., Maier, K., Sottong, R. & Staab, T. E. M. (2011). *J. Phys.: Conf. Ser.* **262**, 012030.

Knipling, K. E., Dunand, D. C. & Seidman, D. N. (2006). *MEKU*. **97**, 246–265.

Knipling, K. E., Dunand, D. C. & Seidman, D. N. (2008). *Acta Materialia*. **56**, 1182–1195.

Labat, S., Richard, M.-I., Dupraz, M., Gailhanou, M., Beutier, G., Verdier, M., Mastropietro, F., Cornelius, T. W., Schülli, T. U., Eymery, J. & Thomas, O. (2015). *ACS Nano*. **9**, 9210–9216.

Larson, F. R. (1952). *Trans. ASME*. **74**, 765–775.

Lo, K. H., Shek, C. H. & Lai, J. K. L. (2009). *Materials Science and Engineering: R: Reports*. **65**, 39–104.

Luke, D. R. (2004). *Inverse Problems*. **21**, 37–50.

Ma, Y., Jiang, B., Li, C., Wang, Q., Dong, C., Liaw, P. K., Xu, F. & Sun, L. (2017). *Metals*. **7**, 57.

Mandula, O., Elzo Aizarna, M., Eymery, J., Burghammer, M. & Favre-Nicolin, V. (2016). *J Appl Cryst*. **49**, 1842–1848.

Marchesini, S., Chapman, H., Hau-Riege, S., London, R., Szoke, A., He, H., Howells, M., Padmore, H., Rosen, R., Spence, J. & others (2003). *Optics Express*. **11**, 2344–2353.

Marquis, E. A. & Seidman, D. N. (2001). *Acta Materialia*. **49**, 1909–1919.

Miao, J., Charalambous, P., Kirz, J. & Sayre, D. (1999). *Nature*. **400**, 342–344.

Miao, J., Sayre, D. & Chapman, H. N. (1998). *J. Opt. Soc. Am. A, JOSAA*. **15**, 1662–1669.

Miyazaki, T., Imamura, H. & Kozakai, T. (1982). *Materials Science and Engineering*. **54**, 9–15.

Oshima, R., Sugiyama, M. & Fujita, F. E. (1988). *MTA*. **19**, 803–810.

Plimpton, S. (1995). *Journal of Computational Physics*. **117**, 1–19.

Plumlee, S. D. (2014). Handcrafting Chain and Bead Jewelry: Techniques for Creating Dimensional Necklaces and Bracelets Potter/Ten Speed/Harmony/Rodale.

Pyczak, F., Devrient, B., Neuner, F. C. & Mughrabi, H. (2005). *Acta Materialia*. **53**, 3879–3891.

Reed, R. (2006). *The Superalloys: Fundamentals and Applications*. Cambridge: Cambridge University Press.

Robinson, I. & Harder, R. (2009). *Nat Mater*. **8**, 291–298.

Semboshi, S., Sasaki, R., Kaneno, Y. & Takasugi, T. (2019). *Metals*. **9**, 160.

Senkov, O. N., Isheim, D., Seidman, D. N. & Pilchak, A. L. (2016). *Entropy*. **18**, 102.

Senkov, O. N., Shagiev, M. R., Senkova, S. V. & Miracle, D. B. (2008). *Acta Materialia*. **56**, 3723–3738.

Shun, T.-T. & Du, Y.-C. (2009). *Journal of Alloys and Compounds*. **479**, 157–160.

Song, G., Sun, Z., Li, L., Xu, X., Rawlings, M., Liebscher, C. H., Clausen, B., Poplawsky, J., Leonard, D. N., Huang, S., Teng, Z., Liu, C. T., Asta, M. D., Gao, Y., Dunand, D. C., Ghosh, G., Chen, M., Fine, M. E. & Liaw, P. K. (2015). *Sci Rep*. **5**, 1–14.

Song, G., Sun, Z., Poplawsky, J. D., Gao, Y. & Liaw, P. K. (2017). *Acta Materialia*. **127**, 1–16.

Stepanov, N. D., Shaysultanov, D. G., Chernichenko, R. S., Tikhonovsky, M. A. & Zherebtsov, S. V. (2019). *Journal of Alloys and Compounds*. **770**, 194–203.

Stukowski, A. (2009). *Modelling Simul. Mater. Sci. Eng.* **18**, 015012.

Teng, Z. K., Miller, M. K., Ghosh, G., Liu, C. T., Huang, S., Russell, K. F., Fine, M. E. & Liaw, P. K. (2010). *Scripta Materialia*. **63**, 61–64.

Thompson, M. E., Su, C. S. & Voorhees, P. W. (1994). *Acta Metallurgica et Materialia*. **42**, 2107–2122.

Tong, C.-J., Chen, Y.-L., Yeh, J.-W., Lin, S.-J., Chen, S.-K., Shun, T.-T., Tsau, C.-H. & Chang, S.-Y. (2005). *Metall and Mat Trans A*. **36**, 881–893.

Tzitzios, V. K., Petridis, D., Zafiropoulou, I., Hadjipanayis, G. & Niarchos, D. (2005). *Journal of Magnetism and Magnetic Materials*. **294**, e95–e98.

Van Sluytman, J. S. & Pollock, T. M. (2012). *Acta Materialia*. **60**, 1771–1783.

Vo, N. Q., Liebscher, C. H., Rawlings, M. J. S., Asta, M. & Dunand, D. C. (2014). *Acta Materialia*. **71**, 89–99.

Volkov, A. Yu. (2004). *Gold Bull*. **37**, 208–215.

Voorhees, P. W. (1992). *Annual Review of Materials Science*. **22**, 197–215.

Wang, Q., Li, Z., Pang, S., Li, X., Dong, C. & Liaw, P. K. (2018). *Entropy*. **20**, 878.

Wang, Q., Ma, Y., Jiang, B., Li, X., Shi, Y., Dong, C. & Liaw, P. K. (2016). *Scripta Materialia*. **120**, 85–89.

Wang, R. W. & Mills, D. L. (1992). *Phys. Rev. B*. **46**, 11681–11687.

Wang, X. G., Liu, J. L., Jin, T. & Sun, X. F. (2014). *Materials & Design*. **63**, 286–293.

Warren, B. E. (1969). *X-Ray Diffraction.* Addison-Wesley, Reading, UK

Wen, S. P., Gao, K. Y., Huang, H., Wang, W. & Nie, Z. R. (2013). *Journal of Alloys and Compounds*. **574**, 92–97.

Xu, W., Birbilis, N., Sha, G., Wang, Y., Daniels, J. E., Xiao, Y. & Ferry, M. (2015). *Nature Mater*. **14**, 1229–1235.

Yau, A., Cha, W., Kanan, M. W., Stephenson, G. B. & Ulvestad, A. (2017). *Science*. **356**, 739–742.

Zhang, L., Zhou, D. & Li, B. (2018). *Materials Letters*. **216**, 252–255.

Zhou, H., Ro, Y., Koizumi, Y., Kobayashi, T., Harada, H. & Okada, I. (2004). *Metall and Mat Trans A*. **35**, 1779–1787.

Zhou, X. W., Johnson, R. A. & Wadley, H. N. G. (2004). *Phys. Rev. B*. **69**, 144113.

Zhou, Y., Liu, Y., Zhou, X., Liu, C., Yu, J., Huang, Y., Li, H. & Li, W. (2017). *Journal of Materials Science & Technology*. **33**, 1448–1456.